\documentclass[%
 reprint,
 superscriptaddress,
 showpacs,
 amsmath,amssymb,
pra,
]{revtex4-1}

\usepackage{graphicx}
\usepackage{dcolumn}
\usepackage{bm}
\usepackage{subcaption}
\usepackage{tabularx,multirow,array,threeparttable}
\usepackage{color}
\usepackage[normalem]{ulem}
\usepackage[
colorlinks=true,
citecolor=blue,
linkcolor=blue,
urlcolor=blue]{hyperref}

\newcommand{\eref}[1]{Eq.\,(\ref{#1})}
\newcommand{\figref}[1]{Fig.\,\ref{#1}}

\begin{document}

\title{
Atomistic mechanism of graphene growth on SiC substrate: Large-scale molecular dynamics simulation based on a new charge-transfer bond-order type potential
}

\author{So Takamoto}
\altaffiliation{Corresponding author}
\email{takamoto.so@fml.t.u-tokyo.ac.jp}
\affiliation{Department of Mechanical Engineering, The University of Tokyo, 7-3-1 Hongo, Bunkyo-ku, Tokyo 113-8656, Japan}
\author{Takahiro Yamasaki}
\affiliation{International Center for Materials Nanoarchitectonics, National Institute for Materials Science, 1-1 Namiki, Tsukuba, Ibaraki 305-0044, Japan}
\author{Jun Nara}
\affiliation{International Center for Materials Nanoarchitectonics, National Institute for Materials Science, 1-1 Namiki, Tsukuba, Ibaraki 305-0044, Japan}
\author{Takahisa Ohno}
\affiliation{International Center for Materials Nanoarchitectonics, National Institute for Materials Science, 1-1 Namiki, Tsukuba, Ibaraki 305-0044, Japan}
\author{Chioko Kaneta}
\affiliation{Fujitsu Laboratories Ltd., 10-1 Morinosato Wakamiya, Atsugi, Kanagawa 243-0197, Japan}
\author{Asuka Hatano}
\affiliation{Department of Mechanical Engineering, The University of Tokyo, 7-3-1 Hongo, Bunkyo-ku, Tokyo 113-8656, Japan}
\author{Satoshi Izumi}
\affiliation{Department of Mechanical Engineering, The University of Tokyo, 7-3-1 Hongo, Bunkyo-ku, Tokyo 113-8656, Japan}

\date{\today}

\begin{abstract}
Thermal decomposition of silicon carbide is a promising approach for the fabrication of graphene.
However, the atomistic growth mechanism of graphene remains unclear.
This paper describes the development of a new charge-transfer interatomic potential.
Carbon bonds with a wide variety of characteristics can be reproduced by the proposed vectorized bond-order term.
Large-scale thermal decomposition simulation enables us to observe the continuous growth process of the multi-ring carbon structure.
The annealing simulation reveals the atomistic process by which the multi-ring carbon structure is transformed to flat graphene involving only 6-membered rings.
Also, it is found that the surface atoms of the silicon carbide substrate enhance the homogeneous graphene formation.

\end{abstract}


\maketitle

\section{Introduction}

Graphene, a single-layer sheet of carbon (C) atoms, has many interesting characteristics \cite{Raimond}, including electronic and mechanical properties.
Graphene is thus expected to have many applications for electronic devices \cite{Lin2010, Morozov2008, Bolotin2008}.
One of the promising methods for the fabrication of graphene is the thermal decomposition of silicon carbide (SiC) \cite{Berger2006, Hibino2008, Virojanadara2008}, because this method enables the graphene to be grown directly on a semi-insulating substrate without the need of a transfer process, which can induce defects in graphene.
In order to obtain high quality and well-controlled graphene sheets, understanding the atomistic growth mechanism is essential.

The characteristics of the graphene depend on the surface orientation of SiC.
On the Si-face of SiC, it is known that homogeneous, monolayer or multilayer graphene can be grown by controlling the growth conditions \cite{Norimatsu2009, Virojanadara2008, Emtsev2009}.
In this paper, we focus on the growth mechanics on the Si-face.

On the Si-face, the step of SiC plays an important role in the graphene growth mechanics \cite{HibinoH.KageshimaH.&Nagase2010, Hannon2008}.
HRTEM observations \cite{Norimatsu2010, Norimatsu2014a} indicate that the initial stage of graphene is nucleated at the facet and grows to the terrace.
Much of the experimental data \cite{Chen2005, Tsai1992, Seyller2006, Charrier2002, Rollings2006} shows that the graphene grown on the Si-face forms a honeycomb lattice consisting of $13\times 13$ graphene rings atop a $6\sqrt{3}\times 6\sqrt{3}\mathrm{R}30$ SiC supercell.
DFT calculations \cite{Varchon2008} indicate that the bottom layer of the graphene has a strong interaction with the SiC substrate.
This is called the 0th layer or buffer layer.

To investigate the atomistic process of graphene growth, several studies with atomic-scale simulations have been performed \cite{Tetlow2014, Norimatsu2014}.
A density functional theory (DFT) study based on geometrical optimization \cite{Kageshima2011, Kageshima2012} showed that graphene is formed on the SiC surface after C atoms are sufficiently accumulated on the surface.
Another study simulated the formation of C clusters \cite{Inoue2012}, and showed that the initial C cluster forms a penta-heptagonal structure (a multi-ring structure with 5- and 7-membered rings) rather than a purely hexagonal structure.
The effect of the step was also investigated.
Static energy calculations for a step model \cite{Kageshima2012} indicated that the aggregation of C atoms at the monolayer step edges of SiC reduces the energy of the system.
Finally, a dynamics simulation in which Si atoms were removed at the facet one-by-one \cite{Morita2013} also suggested that the C multi-ring structure connected to the step edge is created at the nucleation stage.

To the best of our knowledge, all of the existing graphene growth simulation methods are based on the addition of excess C atoms or the removal of pre-selected Si atoms from SiC crystals.
However, the thermal decomposition process is considered to be driven by the recombination of bonds at high temperature, which induces formation of the C cluster and desorption of Si atoms.
In particular, the growing C cluster is considered to be stabilized by connecting to Si atoms \cite{Kageshima2012}.
A previous study \cite{Morita2013} showed that the removal of Si atoms from inappropriate positions has unexpected results.
For example, the desorption of C atoms and the formation of a 3D-shaped C cluster were observed.
Therefore, valid modeling of the desorption of Si atoms is essential in order to observe the graphene growth process until the C cluster completely covers the whole SiC surface.

In addition, the previous reports indicated that the penta-heptagonal structure is created at the initial stage of the graphene nucleation \cite{Inoue2012, Morita2013}.
This structure is considered to be transformed into an ideal graphene structure through the annealing process.
The SiC substrate is thought to strongly affect the annealing process of graphene, since the bottom graphene layer is connected to SiC and has a $6\sqrt{3}\times 6\sqrt{3}\mathrm{R}30$ reconstructed structure.
However, these annealing processes have not yet been investigated by means of atomic-scale simulations.
In order to see the process by which the initial inhomogeneous structure is transformed to ideal graphene through numerous bond recombinations, long-term and large-scale simulations are required.
The DFT calculations in such simulations would have an extremely high computational cost, because of its scope of application.
In this way, there is a large gap between the experiments and simulations.

The classical molecular dynamics (MD) scheme is considered to be an effective approach for these large-scale simulations.
Indeed, several studies have simulated the structure of the graphene on SiC \cite{Lampin2010, Tang2008a} and the graphene growth process on SiC \cite{Tang2008, Jakse2011} using MD.
However, there have been no examples dealing with the continuous growth process of the C cluster, which takes care of the desorption of Si atoms.
In order to reproduce the formation of the C cluster from the SiC substrate through the thermal decomposition using MD, it is necessary for the interatomic potential to express the changes in the nature of the bond accompanying the charge-transfer effect.
This is because Si-C bonds in the SiC substrate have an ionic bond nature due to the difference of electronegativity, while the C cluster has pure covalent bonds.
Also, the C atom shows various bonding nature and is stabilized in various allotropes, e.g., diamond, amorphous carbon and graphite.
The formation of sparse structures such as C chains has also been reported in graphene growth simulations \cite{Morita2013, Ono2016graphene}.
Therefore, the interatomic potential should reproduce the charge-transfer effect and various stable structures of C atoms.
Currently, however, there are no interatomic potentials that meet both requirements.

\section{\label{sec:Method}Methods}

In our previous studies \cite{Takamoto2016, KumagaiSiO2}, we developed a hybrid charge-transfer-type interatomic potential, based on the Tersoff potential, in order to reproduce the thermal oxidation of silicon.
As compared to the other charge-transfer-type interatomic potential \cite{Yu2007, VanDuin2003}, this interatomic potential has the covalent-ionic mixed bond nature as expressed by $f_q$ in \cite{Takamoto2016}.
In this paper, we introduce a new interatomic potential that can reproduce the growth process of graphene on SiC substrate, within the framework of our hybrid charge-transfer interatomic potential.
In addition, we propose an extended potential function without changing the philosophy of the Tersoff potential \cite{Tersoff1986, Tersoff1988a, Tersoff1988b}.
We then incorporate a vectorized environmental-dependent function in order to reproduce various stable structures.

In the Tersoff-type potential, the bond order term $b'$ is calculated as

\begin{equation}
  b'=\left(1+\zeta^n\right)^{-1/(2\sigma)},
\end{equation}
where $n$ and $\sigma$ are potential parameters.
$\zeta$ is a many-body term which expresses the dependence of bond order on the coordination number and bond angle.
In the Tersoff-type potential, $\zeta_{ij}$ ($\zeta$ for bond $i-j$) is calculated as the summation of a function of atoms $i, j$ and $k$ with respect to the surrounding atoms $k$, as shown in \eref{eq:zeta_express}.
The form is

\begin{equation}
  \zeta_{ij}=\sum_{k\neq i,j} F\left(r_{ij}, r_{ik}, \theta_{ijk}\right), \label{eq:zeta_express}
\end{equation}
where F is a function, $r_{ij}$, $r_{ik}$ and $\theta_{ijk}$ are the bond lengths and angles of atoms $i, j$ and $k$.
Here, we call $\zeta$ a surrounding environment function.

In this paper, we have vectorized the surrounding environment function $\zeta$.
In other words, we have defined the bond order $b$ as a function of the independent terms $\zeta_1, \zeta_2, \ldots, \zeta_L$.
Each $\zeta_l$ ($1\leq l\leq L$) is expected to express the respective bonding nature.
For example, $\zeta_1$ corresponds to the structure with a high-coordination number, while $\zeta_2$ corresponds to that with a low-coordination number.

In this paper, the bond order $b$ is defined by the following forms:

\begin{align}
  b&=G\left(\hat{\zeta}_{total}\right)^{-1/(2\sigma)}, \label{eq:vec_bo} \\
  \hat{\zeta}_{total}&=\left(\sum_{l=1}^L\hat{\zeta_l}^{-p}\right)^{-1/p}, \label{eq:hat_zeta} \\
  \hat{\zeta_l}&=g_l+\zeta_l^{n_l},
\end{align}
where $\sigma$, p, $g_l$ and $n_l$ are potential parameters which take positive values.
$G$ is a normalization constant so that $b$ is equal to $1$ in the case that there are no surrounding atoms (which means $\zeta_l =0$ for all $l$). 
Equation \ref{eq:hat_zeta} means ${\hat{\zeta}_{total}}^{-1}$ is calculated as the $L^p$-norm of  ${\zeta_1}^{-1}, {\zeta_2}^{-1}, \ldots, {\zeta_L}^{-1}$.
In the extreme case of $p\rightarrow\infty$, $\hat{\zeta}_{total}$ is equal to the minimum value among $\hat{\zeta_l}$ and other $\hat{\zeta_l}$ are ignored.
In this case, $b$ is rewritten as

\begin{equation}
\begin{split}
  b&=G\left(\min_l\hat{\zeta_l}\right)^{-1/(2\sigma)} \\
  &=\max_l\left\{G\hat{\zeta_l}^{-1/(2\sigma)}\right\} \\
  &=\max_l\left\{G\left(g_l+\zeta_l^{n_l}\right)^{-1/(2\sigma)}\right\}. \label{eq:vec_bo_min}
\end{split}
\end{equation}

Then, we obtained the original Tersoff-type bond order form in the $\max\left\{\;\right\}$ braket.
That means the interatomic potential can select the most stable bonding nature among various $\zeta_l$.
In the case of $p=1$, on the other hand, $\hat{\zeta}_{total}$ becomes the geometric mean, which means that the interatomic potential shows a mixed bond nature.

The entire potential function form is shown in our previous paper \cite{Takamoto2016}.
In this paper, the bond order $b_{ij}$ is calculated by the newly introduced function (\eref{eq:vec_bo}).
We fix $L$ to 2.
We show the function in the expanded form below.

\begin{equation}
	b_{ij}=G\left[\left\{g_1+{({\zeta_1}_{ij})}^{n_1}\right\}^{-p}+\left\{g_2+{({\zeta_2}_{ij})}^{n_2}\right\}^{-p}\right]^{1/2\sigma p},
\end{equation}
where $g_1$, $g_2$, $n_1$, $n_2$, $\delta$ and $p$ are the potential parameters.
$G$ is a normalization constant so that the maximum of $b_{ij}$ will be 1.
In this work, we fix $p$ to 2.
The surrounding environment function ${\zeta_l}_{ij}$ is calculated in the same way as $\zeta_{ij}$ of the previous paper \cite{Takamoto2016}.

The atomic charge $q_i$ varies in response to its environment so as to minimize the total energy $E_{tot}$.
The conjugate gradient method is used to minimize the energy.
The charge is minimized in every timestep.
In addition, the constant energy term is added to the self energy term $U_i^{Self}$ to improve the energy representations of various structures.
This interatomic potential is implemented in the LAMMPS Molecular Dynamics Simulator \cite{plimpton1995fast, lammpssource}.

We fit the potential parameters by our potential-making scheme \cite{Takamoto2016, KumagaiFitting}.
We use various snapshots obtained by MD as a training dataset for fitting.
DFT calculations were performed using the first-principles electronic structure calculation program ``PHASE/0'' \cite{nimsazuma, Ohno:2007:FCL:1362622.1362699}.
We used the local density approximation (LDA).
We typically used $3\times3\times3$ Monkhorst-Pack k-point grids.
The cutoff of plane-wave basis set is 980 eV/atom.
The number of fitted properties is about 260,000.
The properties used for the fitting are energies, atomic forces and atomic charges.

For C cluster fitting, snapshots of diamond, amorphous, liquid, graphene, graphite, and multi-rings with 6-membered rings, single rings with various numbers of atoms and chain structures are sampled. The mixed structures consisting of diamond, 6-membered rings and chains are also sampled.
The same structures are used for Si fitting.
For Si-C system fitting, the following structures were prepared: various SiC crystals, diamond-like structures in which the Si and C atoms are assigned to each site of the diamond lattice in various patterns, diamond-like structures with vacancies, diamond-like structures with surfaces, C clusters on the surface of SiC crystals, flat graphene-like SiC structures and chain structures.
DFT-MD calculations were also performed to sample the structrures of C clusters on the SiC crystals.

\section{\label{sec:Potential}Representation of the interatomic potential}

\begin{figure}
	\centering
	\includegraphics[width=1.\linewidth,trim=20 00 20 00]{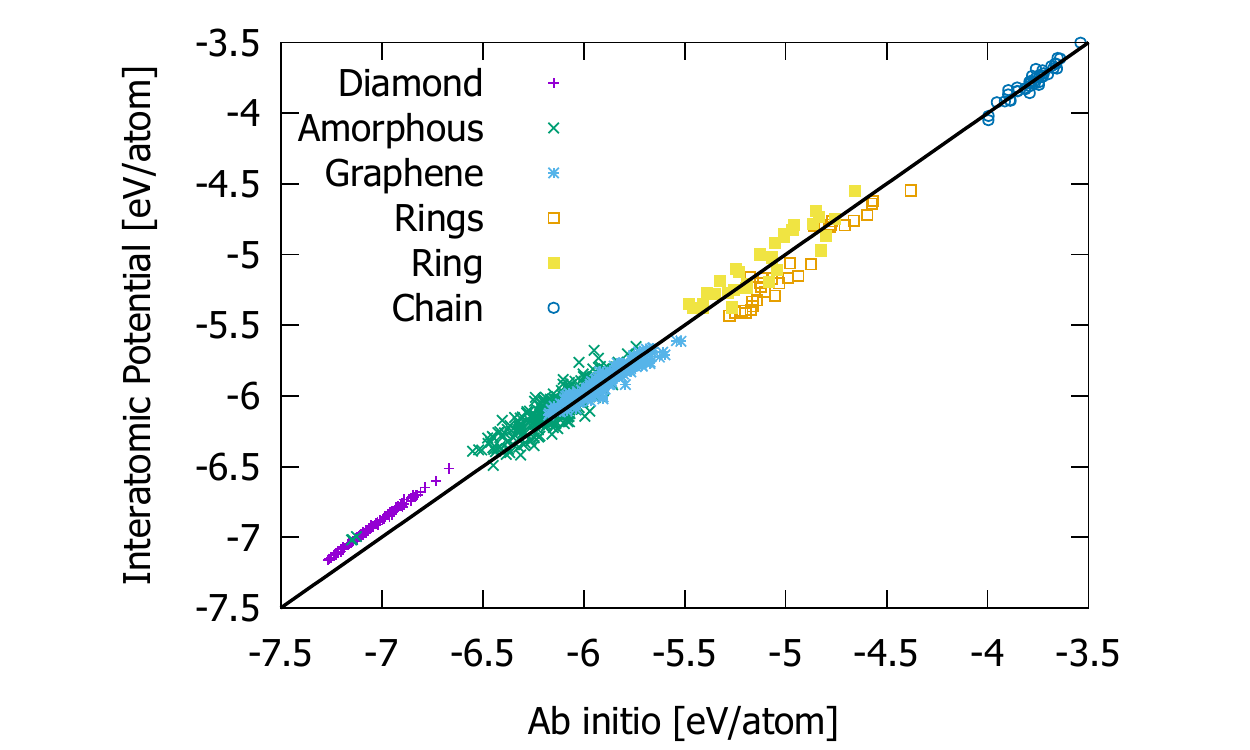}
	\caption{Comparison of the energy representation.}
	\label{fig:fit_c}
\end{figure}

Figure \ref{fig:fit_c} plots the comparison of energies between DFT calculation and the fitted interatomic potential for snapshots of typical structures (diamond, amorphous, graphene, multi-ring, single ring, chain) used for the fitting process.
The fitted interatomic potential well reproduces the structures with a wide-range of coordination numbers.

\begin{figure}
	\begin{minipage}[t]{.49\linewidth}
		\centering
		\includegraphics[width=1.\linewidth,trim=70 00 70 00]{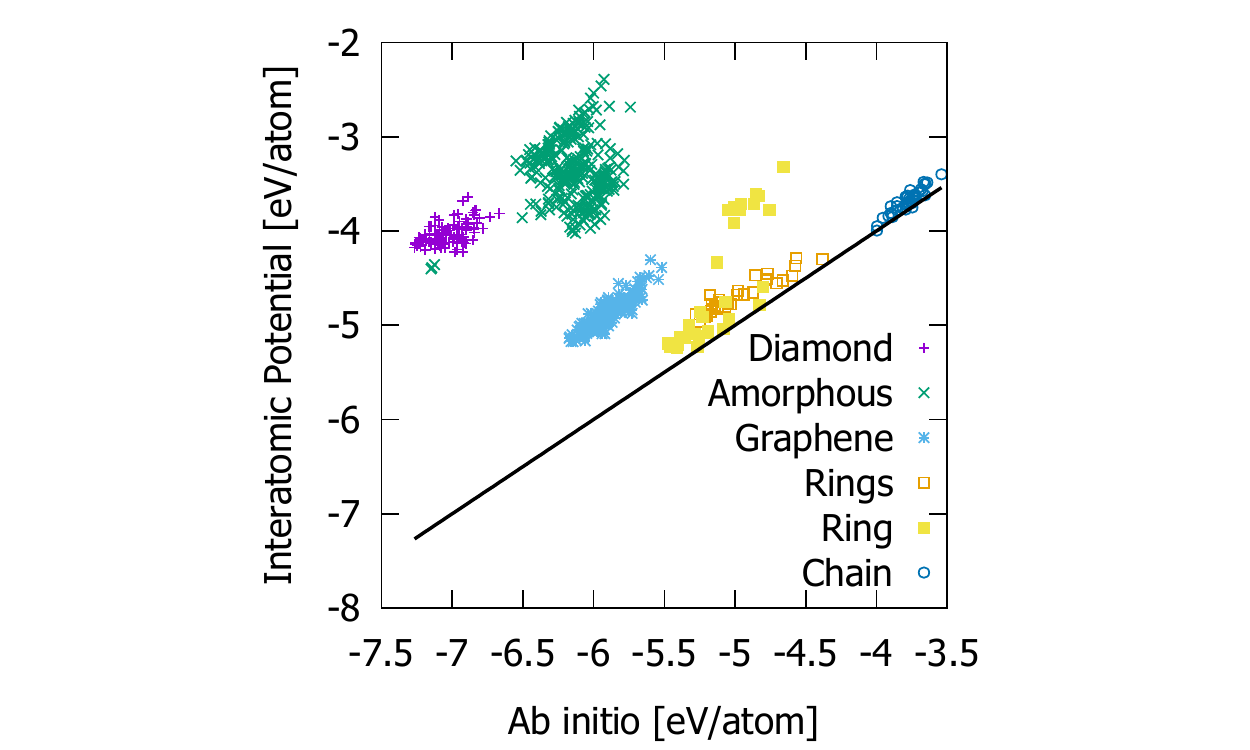}
		\subcaption{Interatomic potential with $\zeta_1$ and suppressed $\zeta_2$.}
		\label{fig:c_k1_1}
	\end{minipage}
	\begin{minipage}[t]{.49\linewidth}
		\centering
		\includegraphics[width=1.\linewidth,trim=70 00 70 00]{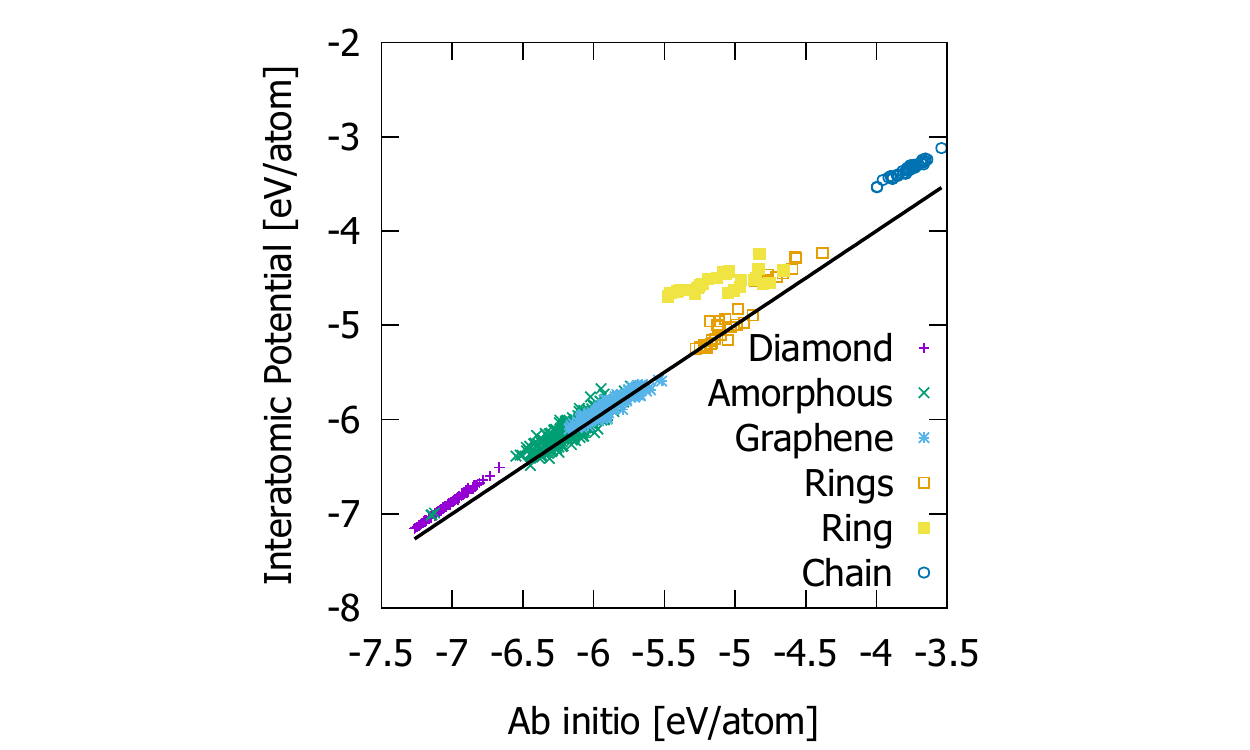}
		\subcaption{Interatomic potential with $\zeta_2$ and suppressed $\zeta_1$.}
		\label{fig:c_k1_2}
	\end{minipage}
	\caption{Energy representation with single bond order.}
	\label{fig:fit_c_k1}
\end{figure}

Next, we verify the effect of the vectorized surrounding environment function $\zeta_l$.
We created interatomic potentials in which one of $\zeta_1, \zeta_2$ is doubled in order to suppress the corresponding bond nature artificially.
The energy comparisons are shown in \figref{fig:fit_c_k1}. 

The interatomic potential in which $\zeta_2$ is suppressed well reproduces the structures with low coordination numbers (e.g., ring and chain), while it has poor energy reproducibility for the structures with high coordination numbers (e.g., diamond and graphene).
On the other hand, the interatomic potential in which $\zeta_1$ is suppressed shows the opposite tendency.
Also, the minimum bond angle of $\zeta_1$ is $180^\circ$ and that of $\zeta_2$ is $127^\circ$.

Next, we verified the reproducibility of sparse structures.
We created a 16-membered single C ring structure and annealed it at 2500 K.
Although the structure of our interatomic potential kept its ring network, that of the Tersoff-type potential \cite{J.Tersoff1994} was transformed into a cage-like high density structure.
The DFT study showed that the ring structure with 16 carbon atoms was more stable than the cage-like structure \cite{Jones1999}.

To validate the applicability of the interatomic potential for the interaction between graphene and SiC, we created a large structure with graphene on the surface of SiC and compared with the previous studies.
As described in the Introduction section, the graphene on the Si-face of SiC is considered to form a $6\sqrt{3}\times 6\sqrt{3}\mathrm{R}30$ structure. We overlaid a $26\times 26$ graphene hexagon structure onto a $12\sqrt{3}\times 12\sqrt{3}$ 4H-SiC supercell.
It is noted that this structure is not used for the fitting process.

\begin{figure}
	\begin{minipage}[t]{1.\linewidth}
		\centering
		\includegraphics[width=1.\linewidth,trim=00 00 00 00]{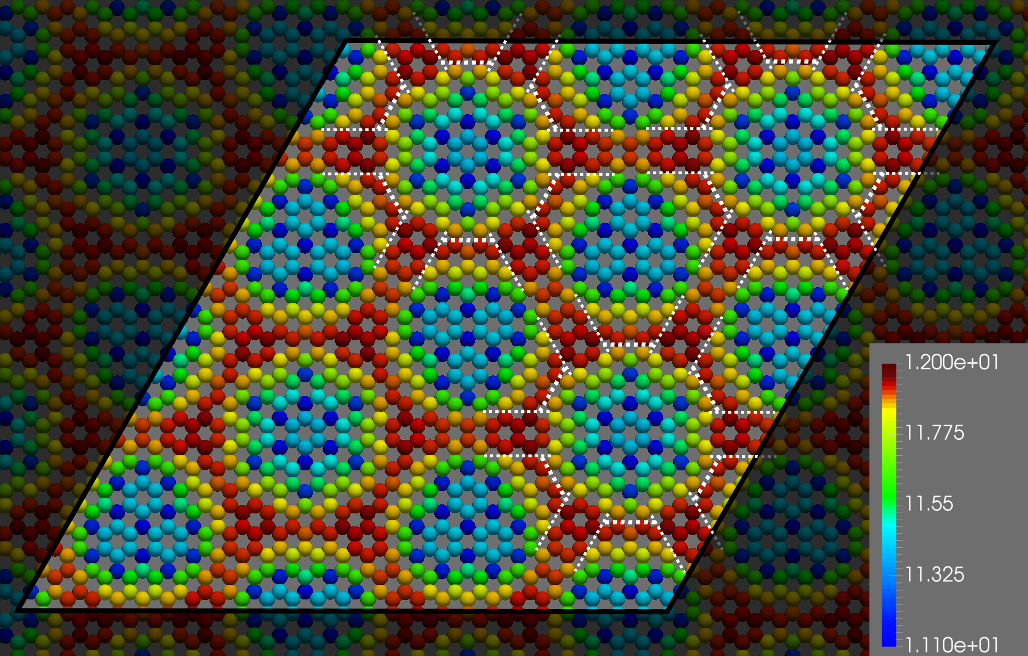}
		\subcaption{Top view.}
		\label{fig:pure_relax_pos}
	\end{minipage}
	\begin{minipage}[t]{1.\linewidth}
		\centering
		\includegraphics[width=1.\linewidth,trim=00 00 00 00]{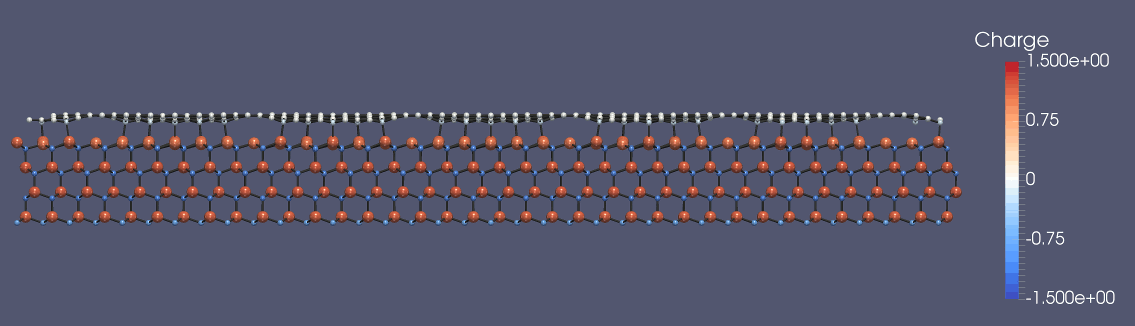}
		\subcaption{Side view.}
		\label{fig:pure_relax_side}
	\end{minipage}
	\caption{Relaxed structure of $6\sqrt{3}\times 6\sqrt{3}\mathrm{R}30$ model. (\ref{fig:pure_relax_pos}): Color of each atom corresponds to the height (length from bottom of the MD cell). Black box corresponds to the MD cell. (\ref{fig:pure_relax_side}): Color of each atom corresponds to the atomic charge.}
	\label{fig:pure_relax}
\end{figure}

The structure after energy minimization is shown in \figref{fig:pure_relax} (top view: (\subref{fig:pure_relax_pos}); side view: (\subref{fig:pure_relax_side})).
The color of \figref{fig:pure_relax}(\subref{fig:pure_relax_pos}) shows the vertical position of the C atoms from the bottom of the MD cell.

From \figref{fig:pure_relax}(\subref{fig:pure_relax_side}), there are two types of C atoms.
One is connected to Si atoms of SiC and the other is not.
This result is consistent with the previous experiment \cite{Chen2005} and DFT \cite{Varchon2008} studies.
The difference in height between the lowest and highest C atoms is about 1.0 $\mathrm{\AA}$.
This value is also close to the DFT result \cite{Varchon2008} (1.2 $\mathrm{\AA}$).
From \figref{fig:pure_relax}(\subref{fig:pure_relax_pos}), we clearly see the long-range hexagonal pattern, which is also seen in the STM image and DFT calculation \cite{Varchon2008}.
As seen by the white dashed line in \figref{fig:pure_relax}(\subref{fig:pure_relax_pos}), there were two types of hexagonal patterns (isotropic ones and triangle ones), which were pointed out in the previous study \cite{Lampin2010}.
Therefore, out interatomic potential reproduces the interactions between SiC and graphene.

\section{\label{sec:Sim}Graphene growth simulation}

\begin{figure}
	\centering
	\includegraphics[width=1.\linewidth,trim=00 00 00 00]{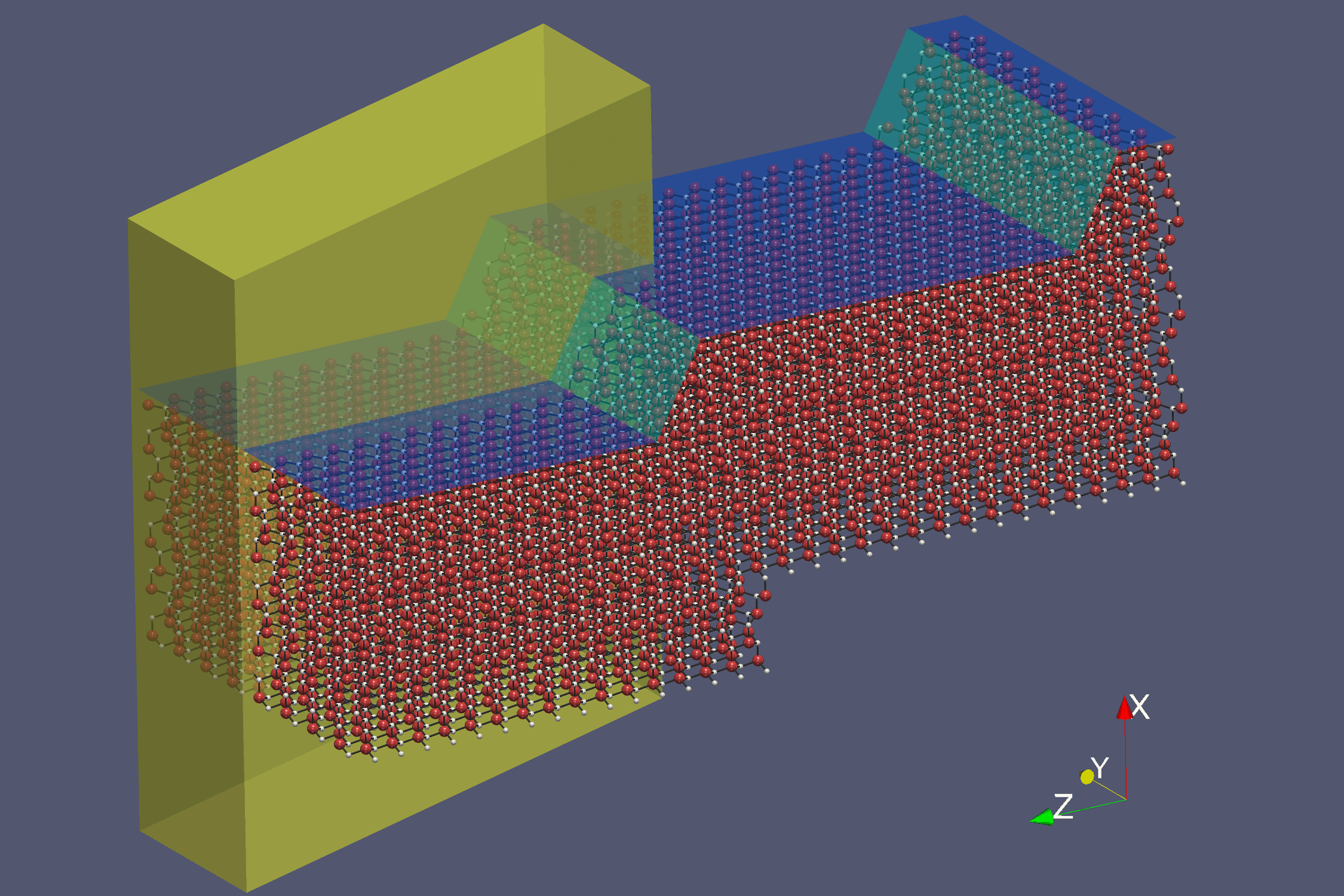}
	\caption{Initial model of graphene growth simulation. Two MD cells are displayed in the Y-axis and the Z-axis direction. The yellow region shows the original MD cell.}
	\label{fig:md_sic_0}
\end{figure}

In order to simulate the dynamics of graphene growth process caused by thermal desorption of Si atoms, we carried out MD simulation.
We have prepared a periodic step structure as shown in \figref{fig:md_sic_0}.
Two MD cells are displayed along the Y and Z directions.
The substrate is 4H-SiC.
The surface orientations of the terrace and facet are $(0001)$ and $(11\overline{2}2)$, respectively.
The terrace and facet faces are shown in blue and green planes.
The height of the facet is one layer of the 4H-SiC lattice (4 bilayer).
As indicated by the yellow box in \figref{fig:md_sic_0}, the MD cell is tilted so that the upper and lower terraces are smoothly connected without any steps through the periodic boundary.
The total number of atoms in the single MD cell is about 12,000.

In this simulation, the desorption of Si atoms is modeled by gradually removing instable Si atoms on the surface during MD.
The stability of Si atoms is measured by the total energy change following the elimination of that atom at every 1 ps.
The threshold of removing Si atoms is set to be 6.5 eV.
In this simulation, the energy to remove a typical Si atom on the terrace is about 10 eV.
Therefore, the Si atoms whose energy is 3.5 eV higher than that of Si atoms on the terrace are removed.
In this manner, the order and timing of Si atoms to be removed are not set in advance.

We carried out MD with NVT ensemble at 2500 K while fixing the bottom of SiC.
It is noted that the experimental temperature for growth process is typically between 1473 K and 1933 K \cite{Tetlow2014}, and the growth timescale is the order of minutes.

\begin{figure*}
	\begin{minipage}[t]{1.\linewidth}
		\centering
		\includegraphics[width=.49\linewidth,trim=00 00 00 00]{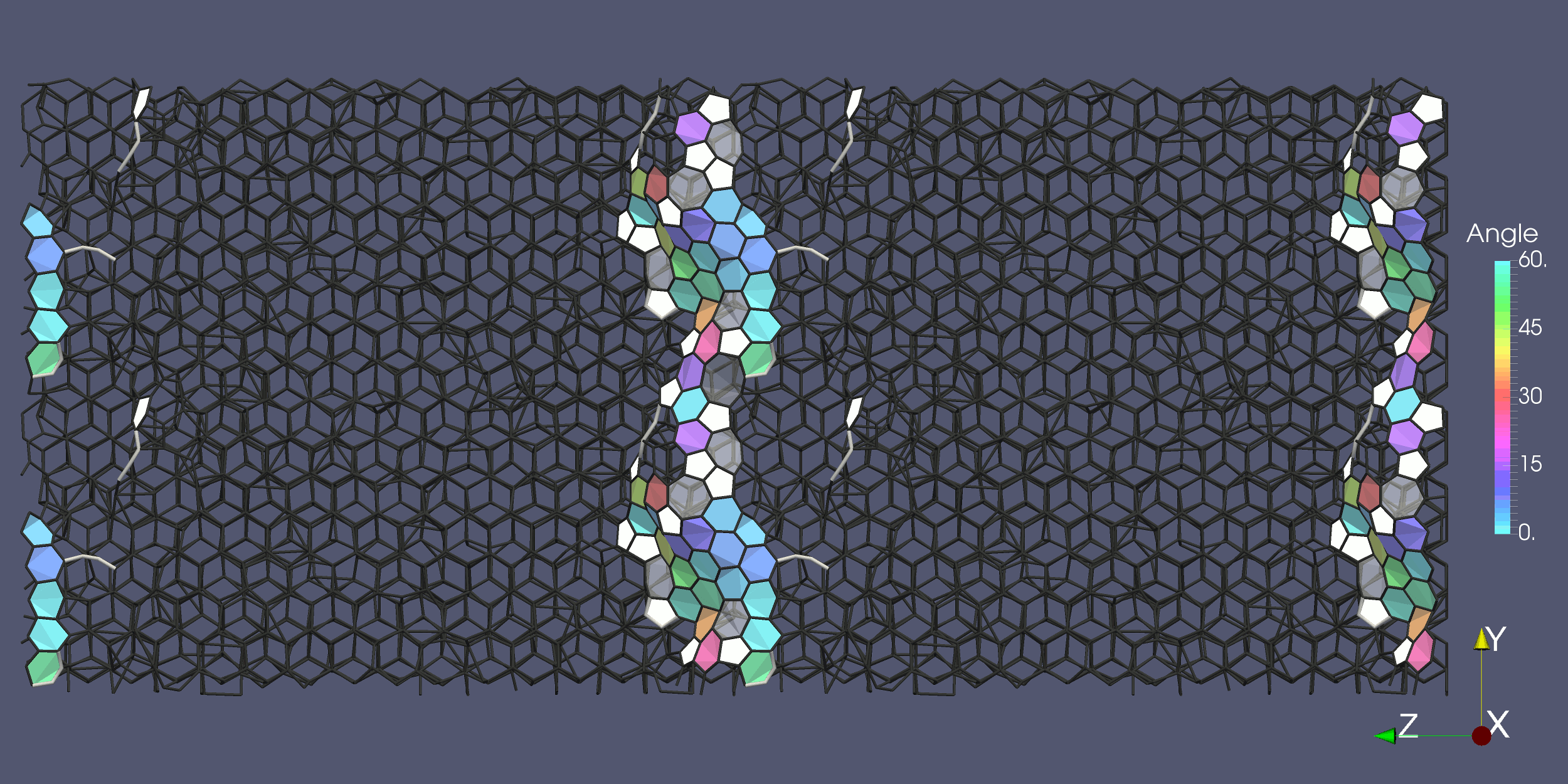}
		\includegraphics[width=.49\linewidth,trim=00 00 00 00]{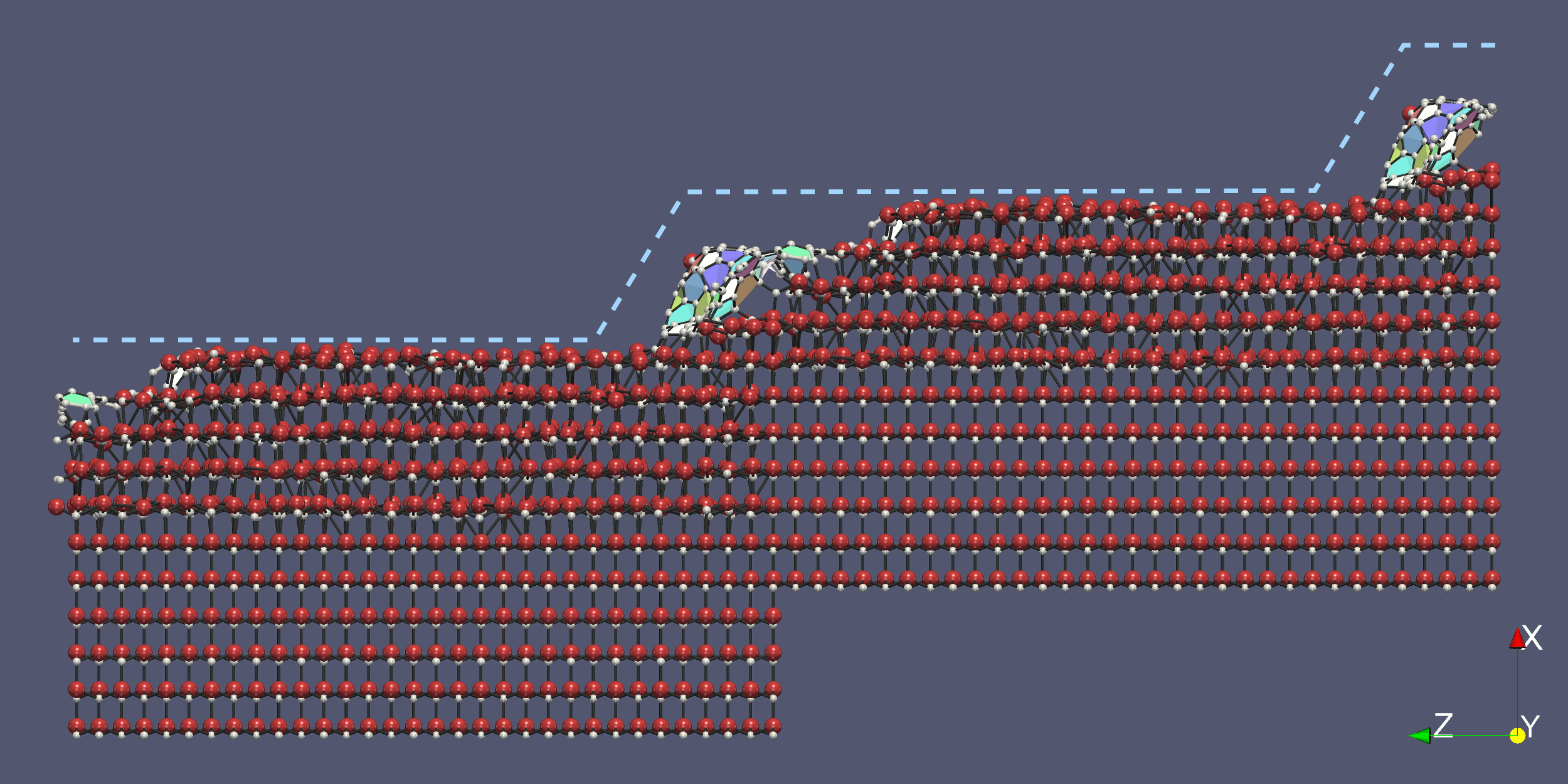}
		\subcaption{0.5 ns.}
		\label{fig:md_sic_side_10}
	\end{minipage}
	\begin{minipage}[t]{1.\linewidth}
		\centering
		\includegraphics[width=.49\linewidth,trim=00 00 00 00]{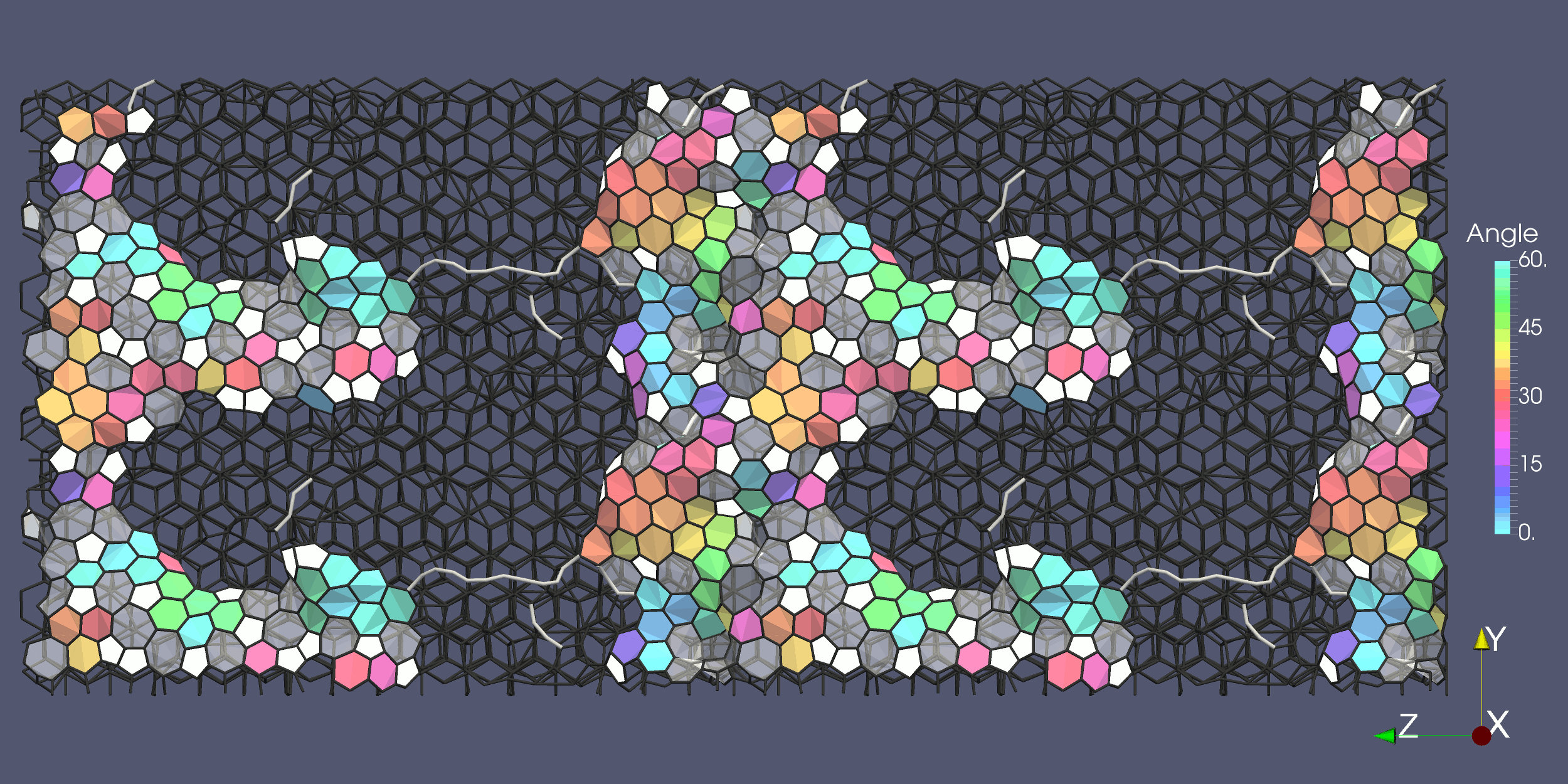}
		\includegraphics[width=.49\linewidth,trim=00 00 00 00]{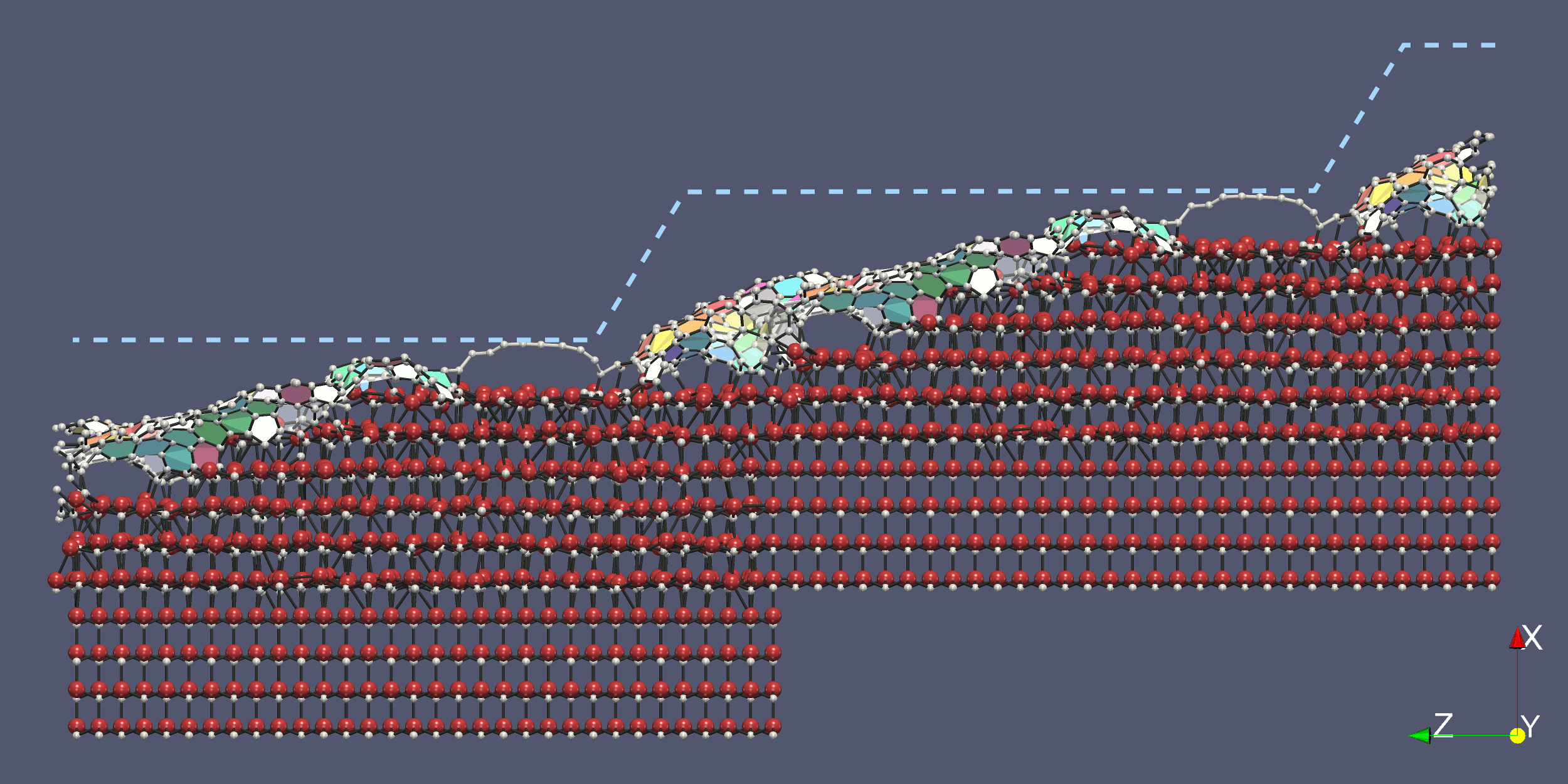}
		\subcaption{1.6 ns.}
		\label{fig:md_sic_side_32}
	\end{minipage}
	\begin{minipage}[t]{1.\linewidth}
		\centering
		\includegraphics[width=.49\linewidth,trim=00 00 00 00]{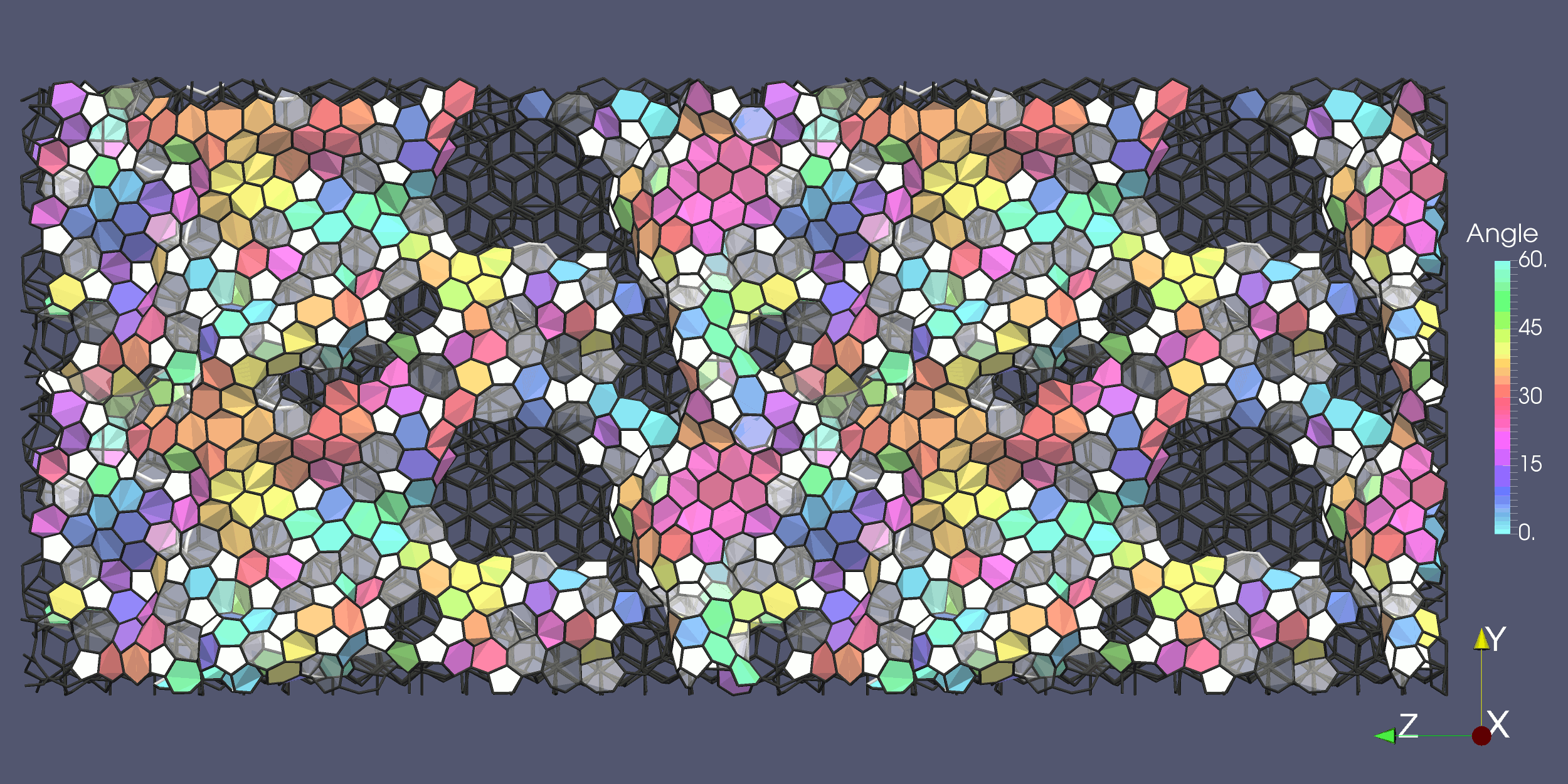}
		\includegraphics[width=.49\linewidth,trim=00 00 00 00]{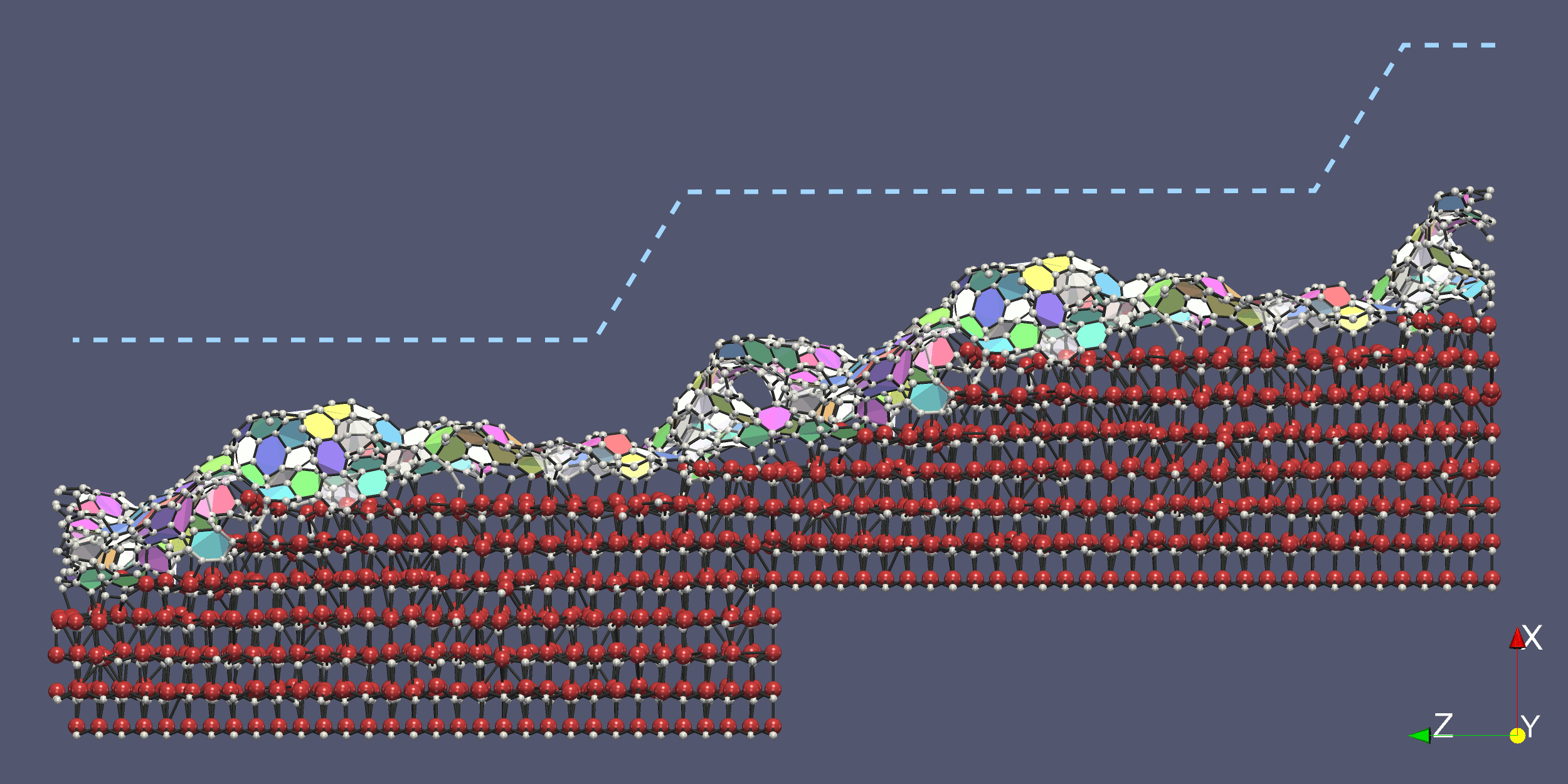}
		\subcaption{3.2 ns.}
		\label{fig:md_sic_side_64}
	\end{minipage}
	\caption{Process of the graphene growth simulation. 6-membered rings are colored according to their orientation. Left: top view. C ring structures, C chains and bonds are shown. Right: side view. Si atoms (larger spheres) and C atoms (smaller spheres) are also shown.}
	\label{fig:md_sic_side}

	\begin{minipage}[t]{1.\linewidth}
		\centering
		\includegraphics[width=.49\linewidth,trim=00 00 00 00]{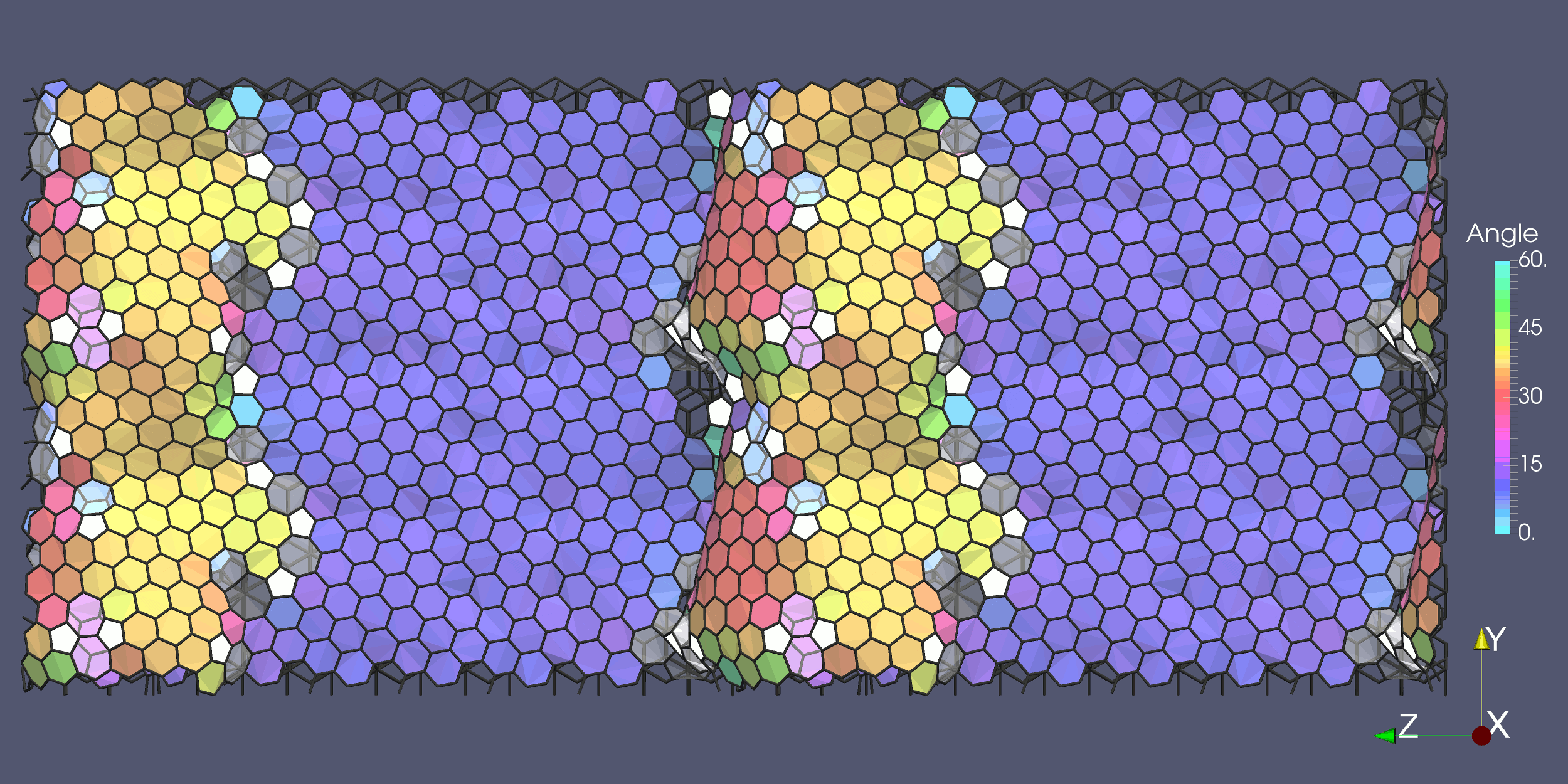}
		\includegraphics[width=.49\linewidth,trim=00 00 00 00]{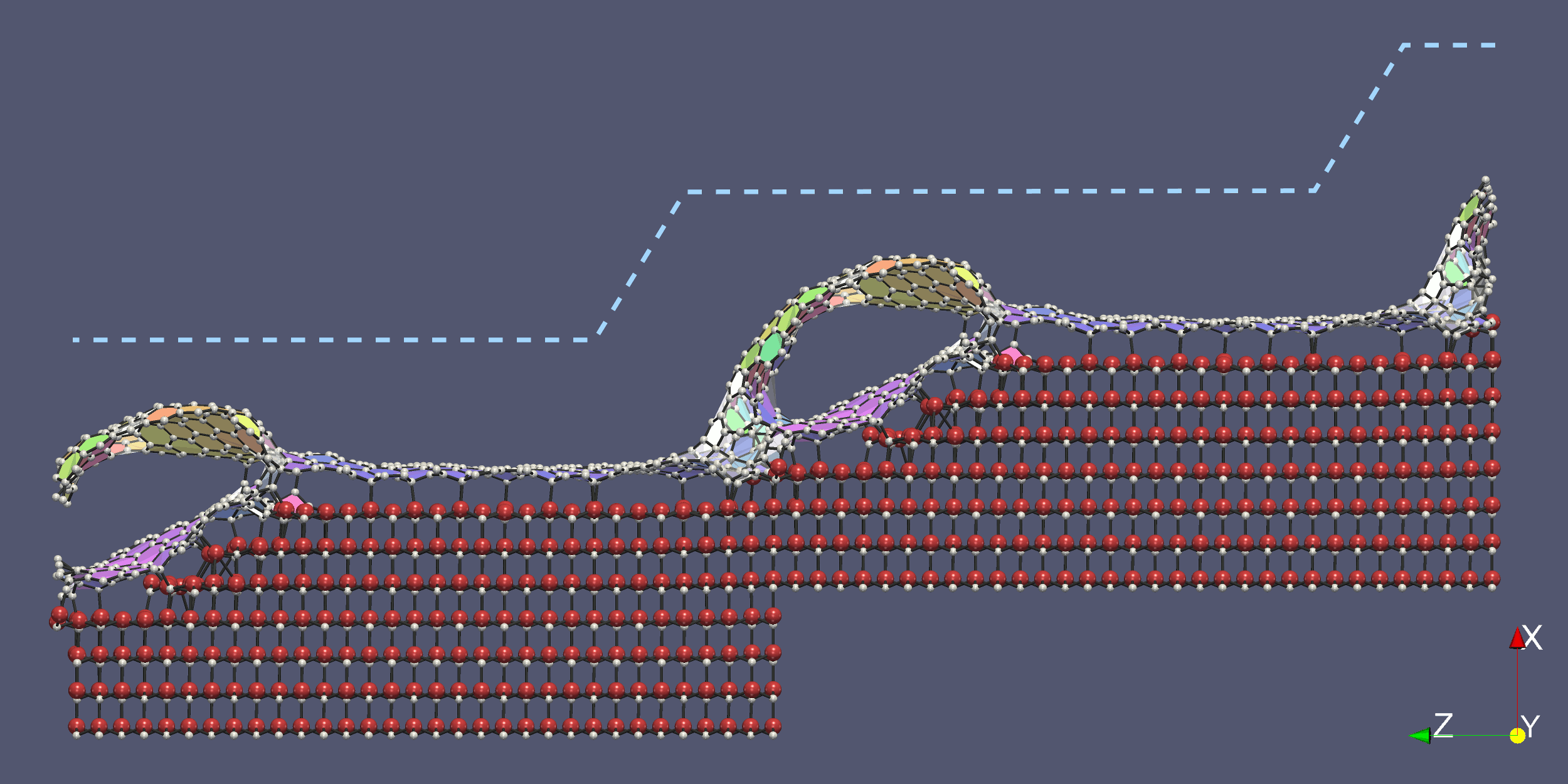}
	\end{minipage}
	\caption{The final structure obtained by the annealing process. After 3.2 ns of thermal decomposition simulation, the structure was annealed for 12.8 ns and cooled to 0 K. Left: top view. Right: side view.}
	\label{fig:md_sic_anneal}
\end{figure*}

The results are shown in \figref{fig:md_sic_side}.
Here, we illustrate C rings as films.
7 or 8-membered rings are shown in (semitransparent) gray and 3,4 or 5-membered rings are shown in white.
The hue of the 6-membered ring expresses its in-plane orientation.
The hue makes one lap at 60 degrees rotation.
The orientation is zero in the case that one pair of the diagonal vertex is directed to $[11\overline{2}0]$.
We also illustrate C chains as white lines.
The left figures are top views of the rings, chains and bonds.
The right figures are side views in which atoms are shown in addition to the rings, chains and bonds.
The dashed white lines show the initial position of the surface.

The graphene growth proceeded as follows.
Figure \ref{fig:md_sic_side}(\subref{fig:md_sic_side_10}): Desorption of Si atoms from the facet is observed.
As the number of C atoms on the facet increases, a striped multi-ring structure is formed.
The bottom of the C cluster is vertically bonded to the SiC.
While the total number of the rings is less than 20, the numbers of 5-membered rings and 6-membered rings are increased in the same way as in the previous study \cite{Morita2013}.

Figure \ref{fig:md_sic_side}(\subref{fig:md_sic_side_32}): After the formation of the C cluster on the facet face, the desorption of Si atoms at the edge of the upper terrace starts and creates one-bilayer steps.
The excess C atoms move on the terrace actively and connect to the C cluster.
As a result, the C multi-ring structure is formed on the terrace.
C chains connected to the C cluster are also seen on the terrace.
The C multi-ring structure on the terrace is connected to the edge of the step of the SiC bilayer.
It is noted that the top bilayer also consists of 6-membered rings and it can smoothly connect to the C multi-ring structure.
Also, the aggregation of C atoms forms a C-rich area and Si-rich area.
In \figref{fig:md_sic_side}(\subref{fig:md_sic_side_32}), the shape of the C cluster is wavy.
This shape is similar to the finger-like structures of graphene observed by AFM \cite{Borovikov2009}.
However, caution should be taken when making a direct comparison, because this structure has a ten-nanometer scale, while the observed finger-like structures have a micrometer scale.

Figure \ref{fig:md_sic_side}(\subref{fig:md_sic_side_64}): The desorption of Si atoms continues and reaches to the other side of the C cluster.
C chains are wound up by the multi-ring structure.
As a result, a sheet composed of C multi-ring structures covers over the entire surface.

\section{\label{sec:Discussion}Discussion}

\begin{figure}
	\centering
	\includegraphics[width=1.\linewidth,trim=00 00 00 00]{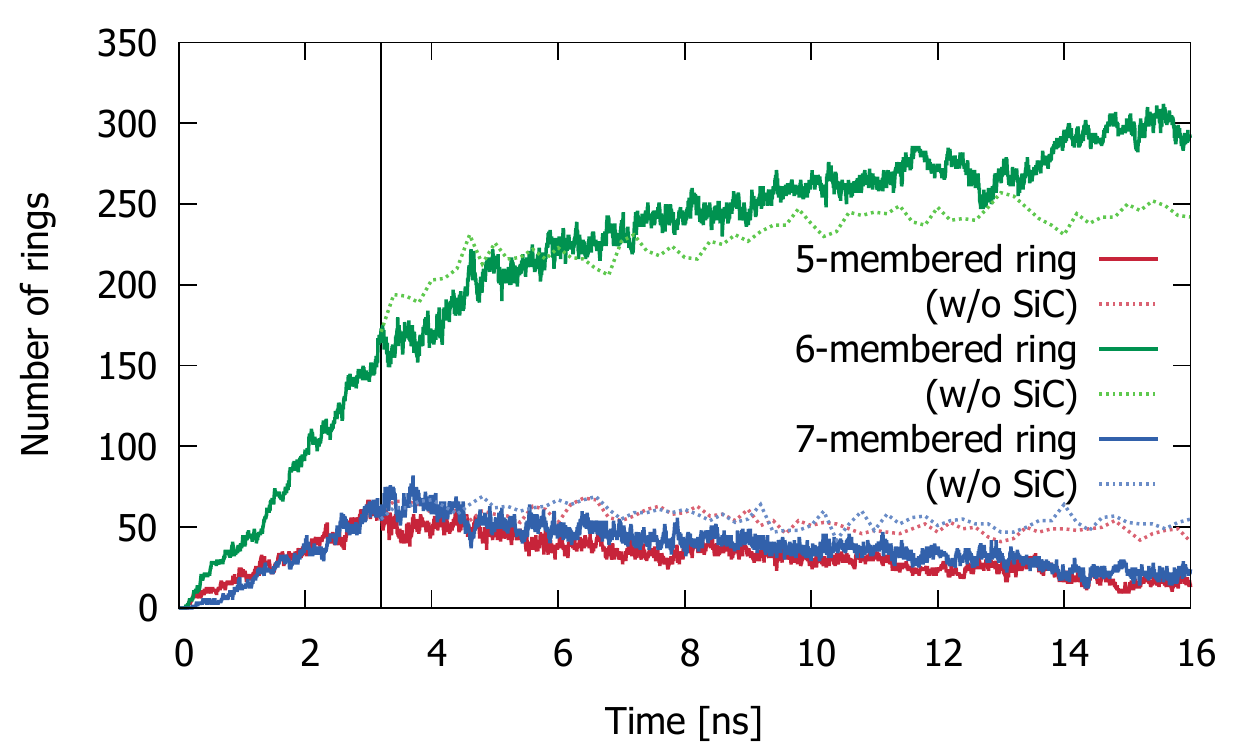}
	\caption{Time histories of the number of rings.}
	\label{fig:md_rings_num}
\end{figure}

The obtained structure involves the roughness due to many 5 and 7-membered rings, which is similar to the previous DFT results \cite{Inoue2012}.
In addition, the structure is inhomogeneous and contains thick areas and holes.
It is considered that the structure is transformed into a perfect graphene structure by the additional annealing process.
Therefore, the structure is annealed at 3000 K NVT ensemble for 12.8 ns.
After the annealing simulation, we relaxed the structure.
The annealed structure is shown in \figref{fig:md_sic_anneal}.
The time histories for the numbers of 5, 6, and 7-membered rings are shown in \figref{fig:md_rings_num}.
The annealing starts at 3.2 ns (vertical line in the figure).

\begin{figure}
	\centering
	\includegraphics[width=1.\linewidth,trim=00 00 00 00]{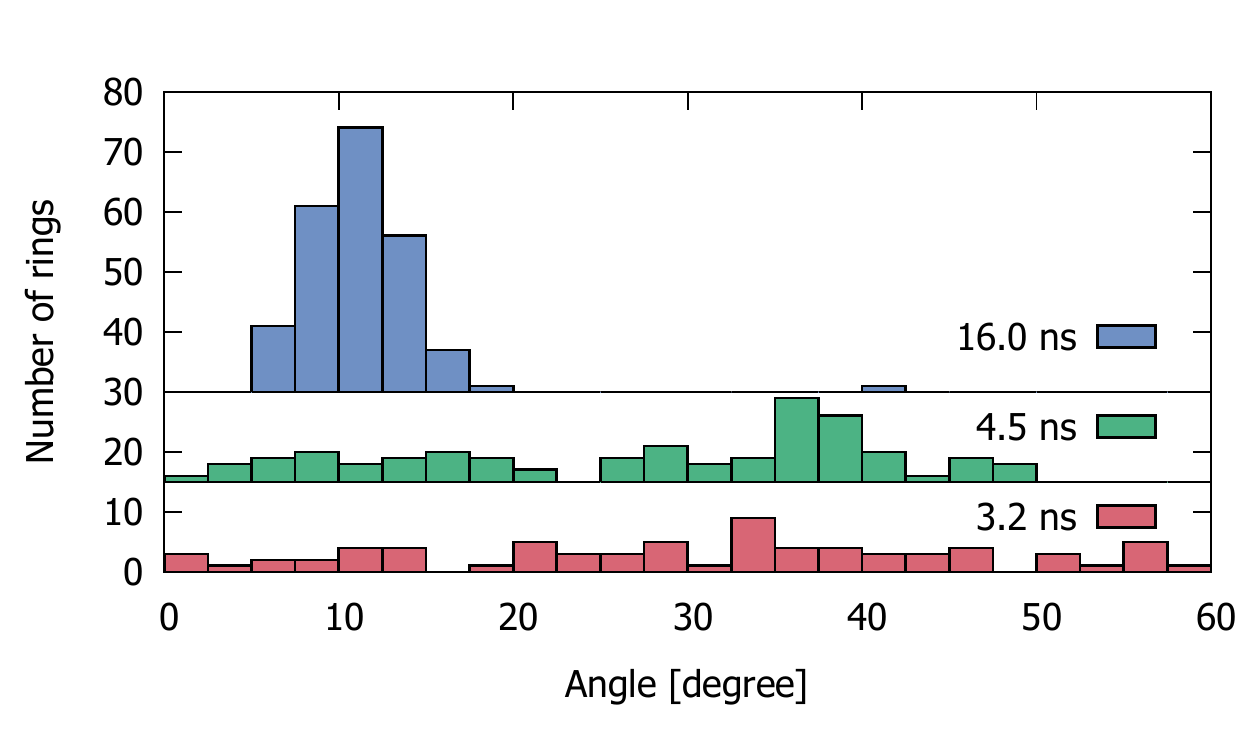}
	\caption{Histogram of the orientations of 6-membered rings on the terrace.}
	\label{fig:md_rings_angle}
\end{figure}

During the annealing, the number of 6-membered rings is increased and those of 5 and 7-membered rings are decreased.
The diffusion of C atoms makes the density of C atoms uniform.
As a result, a flat, graphene-like structure was obtained on the terrace, while a double layer structure was created on the facet due to excess C atoms.
In particular, a perfect and flat graphene structure composed of 6-membered rings was formed on the terrace.

Histograms of the orientations of 6-membered rings on the terrace after 3.2, 4.5 and 16.0 ns are shown in \figref{fig:md_rings_angle}.
Zero degrees corresponds to the $6\sqrt{3}\times 6\sqrt{3}\mathrm{R}30$ structure.
At the initial stage (4.5 ns), the largest graphene piece orients $35^{\circ}$.
However, the further annealing gradually shrinks this orientation region.
The surface atoms of the SiC substrate are likely to affect this process since $35^{\circ}$ deviates from the stable orientation.
The peak of $10^{\circ}$ after 16 ns would be caused by the artificial periodic boundary conditions because it limits the possible orientation and prevents the rotation of the structure.

\begin{figure}
	\centering
	\includegraphics[width=1.\linewidth,trim=00 00 00 00]{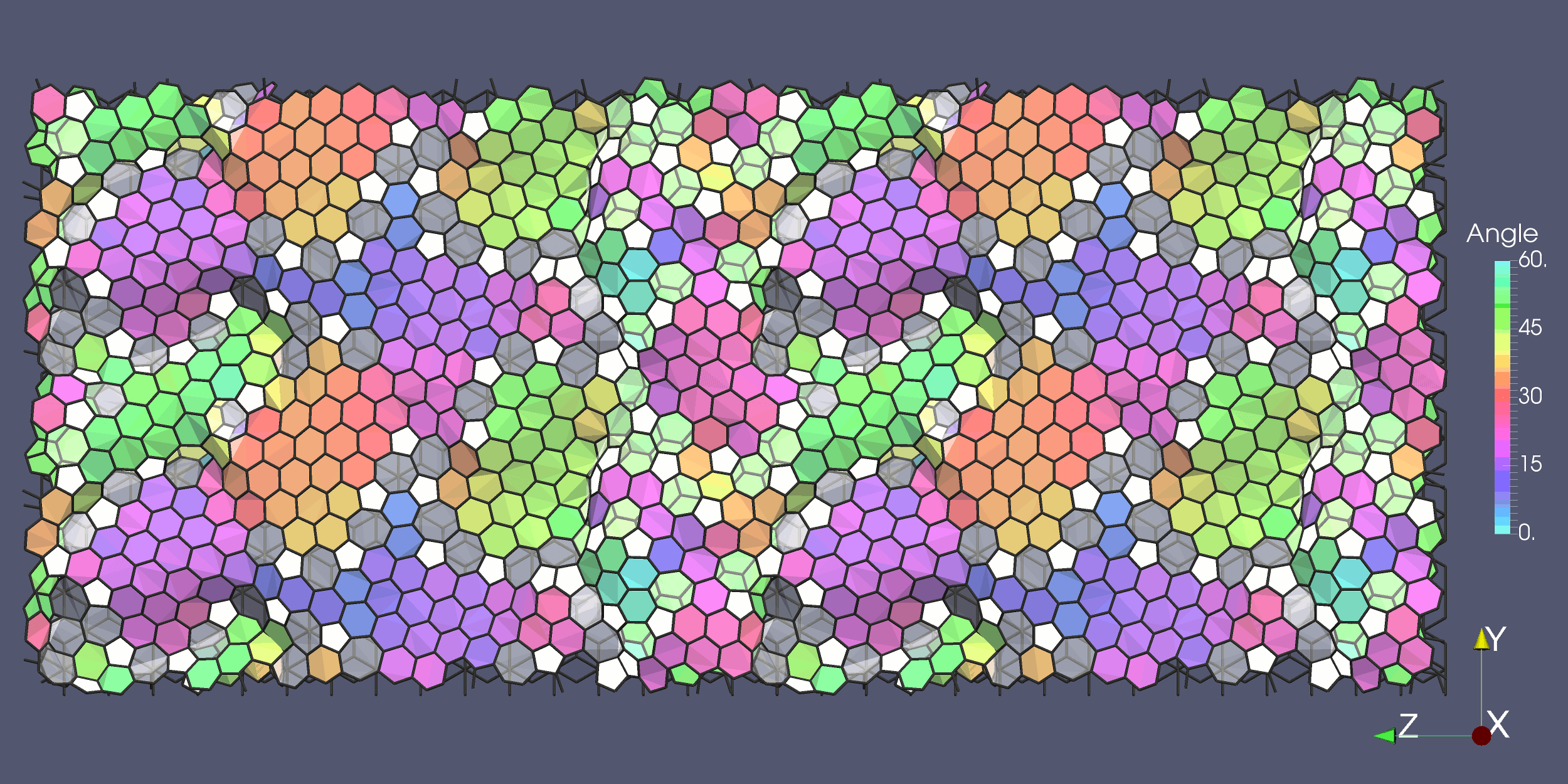}
	\caption{C cluster annealed without SiC.}
	\label{fig:md_conly_anneal}
\end{figure}

In order to investigate the effect of the SiC substrate on the formation of graphene, we created a ``free-standing C cluster film'' model by eliminating SiC from the 3.2 ns structure (\figref{fig:md_sic_side}(\subref{fig:md_sic_side_64})).
After the same-condition annealing, the structure shown in \figref{fig:md_conly_anneal} was obtained.
The time histories for the numbers of 5, 6, and 7-membered rings are also shown in \figref{fig:md_rings_num} (the lines labeled ``w/o SiC'').
It can be seen that the region of 6-membered rings is divided into small pieces and its orientation is not uniform.
During the annealing, the number of 5 and 7-membered rings is decreased by 20 \%, while it is decreased by 80 \% in the case of graphene with SiC.
Therefore, it is considered that the SiC substrate enhances the recombination of C-C bonds and resulting homogeneous formation.
This would be similar to the graphene growth on metals.
The calculated reaction barrier to healing a Stone-Wales defect on an Ni $(111)$ \cite{Jacobson2012} is 2.88 eV, which is lower than that for free-standing graphene (4.10 eV).

\begin{figure}
	\begin{minipage}[t]{1.\linewidth}
		\centering
		\includegraphics[width=.49\linewidth,trim=00 00 00 00]{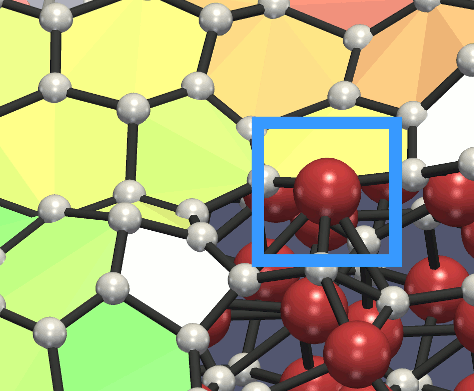}
		\includegraphics[width=.49\linewidth,trim=00 00 00 00]{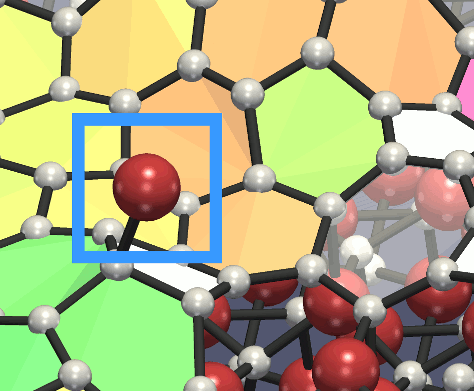}
		\subcaption{Release of Si atom.}
		\label{fig:md_sic_move_pop_before}
	\end{minipage}
	\begin{minipage}[t]{1.\linewidth}
		\centering
		\includegraphics[width=1.\linewidth,trim=00 00 00 00]{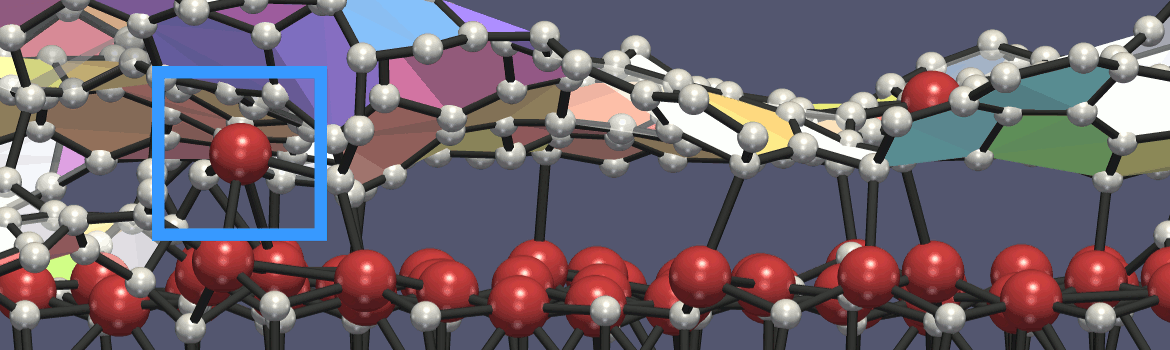}
		\includegraphics[width=1.\linewidth,trim=00 00 00 00]{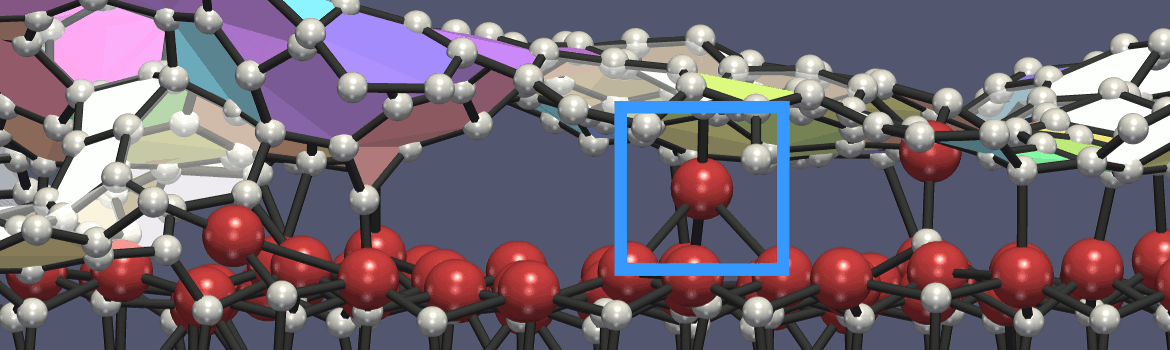}
		\subcaption{Diffusion of Si atom.}
		\label{fig:md_sic_move_diffuse_before}
	\end{minipage}
	\caption{The active movement of Si atoms at the surface and the interface.}
	\label{fig:md_si_move}
\end{figure}

It is noted that Si atoms frequently move as readily as C atoms.
During the annealing simulation, we observed that several Si atoms were ejected to the surface side of the C cluster film through a hole, and were diffused along the SiC-graphene interface.
The snapshots of these events are shown in \figref{fig:md_si_move}.
These results indicate that the thermal decomposition of SiC continues even after the surface is covered by the graphene.

\begin{figure}
	\centering
	\includegraphics[width=1.\linewidth,trim=00 00 00 00]{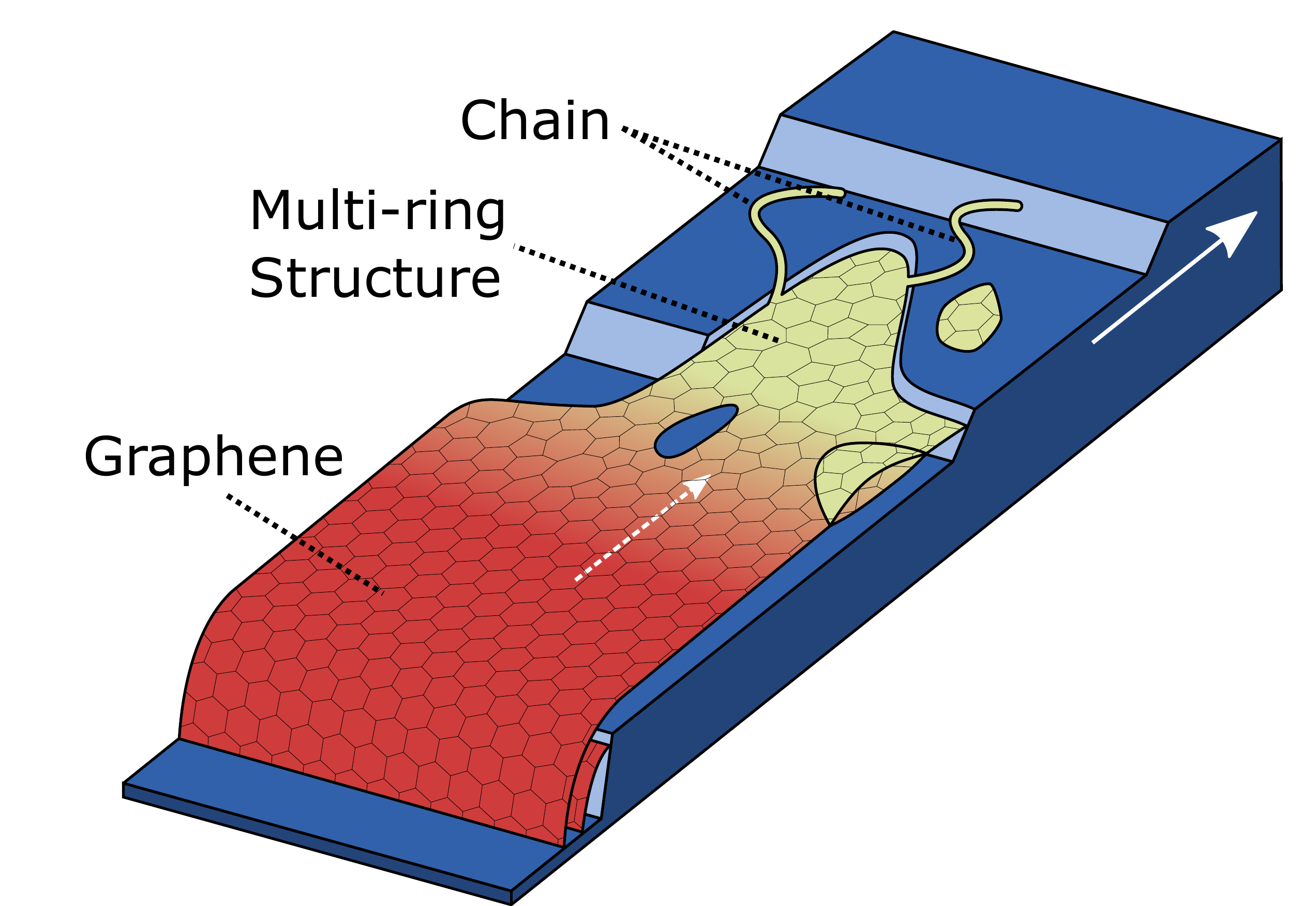}
	\caption{Graphene growth model.}
	\label{fig:md_model}
\end{figure}

The graphene growth model suggested from these results is shown in \figref{fig:md_model}.
After the formation of the C cluster at the facet, the desorption of Si atoms starts at the upper terrace.
The excess C atoms move actively on the SiC and form multi-ring structures.
These processes are consistent with the model inferred by the earlier experimental work \cite{Norimatsu2010}.
The initial C cluster is not uniform and contains many 5 and 7-membered rings.
The SiC surface enhances the bond recombinations of the C cluster and the formation of the perfect graphene.
It should be noted that the graphene growth mode may depend on the direction and height of the step and crystal polymorphism \cite{dojima2016kinetically}.
Investigation of those effects is a future work.

\section{\label{sec:Conclusion}Conclusion}

A new charge-transfer interatomic potential is developed in order to reproduce the formation and homogenization of graphene on the SiC by thermal decomposition.
A new bond order function based on a vectorized surrounding environment parameter is proposed to reproduce various bond nature of C atoms.

The large-scale thermal decomposition simulation reproduces the continuous growth process of the C multi-ring structure on the terrace.
The annealing simulation reveals the atomistic process of the transformation of the C multi-ring structure to flat graphene involving only 6-membered rings.
Also, it is found that the surface atoms of the SiC substrate enhance the homogeneous graphene formation.

\begin{acknowledgments}
A portion of this research was partly supported by MEXT within the priority issue 6 of the FLAGSHIP2020 and JSPS KAKENHI Grant No. 16H03830.

The DFT-MD calculations were carried out on the K computer provided by RIKEN, AICS through the HPCI System Research project (Project ID:hp150266 and hp160226).

This work was supported by a Grant-in-Aid for JSPS Research Fellow.
\end{acknowledgments}

\bibliography{main.bbl}

\begin{thebibliography}{47}%
\makeatletter
\providecommand \@ifxundefined [1]{%
 \@ifx{#1\undefined}
}%
\providecommand \@ifnum [1]{%
 \ifnum #1\expandafter \@firstoftwo
 \else \expandafter \@secondoftwo
 \fi
}%
\providecommand \@ifx [1]{%
 \ifx #1\expandafter \@firstoftwo
 \else \expandafter \@secondoftwo
 \fi
}%
\providecommand \natexlab [1]{#1}%
\providecommand \enquote  [1]{``#1''}%
\providecommand \bibnamefont  [1]{#1}%
\providecommand \bibfnamefont [1]{#1}%
\providecommand \citenamefont [1]{#1}%
\providecommand \href@noop [0]{\@secondoftwo}%
\providecommand \href [0]{\begingroup \@sanitize@url \@href}%
\providecommand \@href[1]{\@@startlink{#1}\@@href}%
\providecommand \@@href[1]{\endgroup#1\@@endlink}%
\providecommand \@sanitize@url [0]{\catcode `\\12\catcode `\$12\catcode
  `\&12\catcode `\#12\catcode `\^12\catcode `\_12\catcode `\%12\relax}%
\providecommand \@@startlink[1]{}%
\providecommand \@@endlink[0]{}%
\providecommand \url  [0]{\begingroup\@sanitize@url \@url }%
\providecommand \@url [1]{\endgroup\@href {#1}{\urlprefix }}%
\providecommand \urlprefix  [0]{URL }%
\providecommand \Eprint [0]{\href }%
\providecommand \doibase [0]{http://dx.doi.org/}%
\providecommand \selectlanguage [0]{\@gobble}%
\providecommand \bibinfo  [0]{\@secondoftwo}%
\providecommand \bibfield  [0]{\@secondoftwo}%
\providecommand \translation [1]{[#1]}%
\providecommand \BibitemOpen [0]{}%
\providecommand \bibitemStop [0]{}%
\providecommand \bibitemNoStop [0]{.\EOS\space}%
\providecommand \EOS [0]{\spacefactor3000\relax}%
\providecommand \BibitemShut  [1]{\csname bibitem#1\endcsname}%
\let\auto@bib@innerbib\@empty
\bibitem [{\citenamefont {Novoselov}\ \emph {et~al.}(2004)\citenamefont
  {Novoselov}, \citenamefont {Geim}, \citenamefont {Morozov}, \citenamefont
  {Jiang}, \citenamefont {Zhang}, \citenamefont {Dubonos}, \citenamefont
  {Grigorieva},\ and\ \citenamefont {Firsov}}]{Raimond}%
  \BibitemOpen
  \bibfield  {author} {\bibinfo {author} {\bibfnamefont {K.~S.}\ \bibnamefont
  {Novoselov}}, \bibinfo {author} {\bibfnamefont {A.~K.}\ \bibnamefont {Geim}},
  \bibinfo {author} {\bibfnamefont {S.~V.}\ \bibnamefont {Morozov}}, \bibinfo
  {author} {\bibfnamefont {D.}~\bibnamefont {Jiang}}, \bibinfo {author}
  {\bibfnamefont {Y.}~\bibnamefont {Zhang}}, \bibinfo {author} {\bibfnamefont
  {S.~V.}\ \bibnamefont {Dubonos}}, \bibinfo {author} {\bibfnamefont {I.~V.}\
  \bibnamefont {Grigorieva}}, \ and\ \bibinfo {author} {\bibfnamefont {a.~a.}\
  \bibnamefont {Firsov}},\ }\href {\doibase 10.1126/science.1102896} {\bibfield
   {journal} {\bibinfo  {journal} {Science (New York, N.Y.)}\ }\textbf
  {\bibinfo {volume} {306}},\ \bibinfo {pages} {666} (\bibinfo {year}
  {2004})},\ \Eprint {http://arxiv.org/abs/0410550} {arXiv:0410550 [cond-mat]}
  \BibitemShut {NoStop}%
\bibitem [{\citenamefont {Lin}\ \emph {et~al.}(2010)\citenamefont {Lin},
  \citenamefont {Dimitrakopoulos}, \citenamefont {Jenkins}, \citenamefont
  {Farmer}, \citenamefont {Chiu}, \citenamefont {Grill},\ and\ \citenamefont
  {Avouris}}]{Lin2010}%
  \BibitemOpen
  \bibfield  {author} {\bibinfo {author} {\bibfnamefont {Y.-M.}\ \bibnamefont
  {Lin}}, \bibinfo {author} {\bibfnamefont {C.}~\bibnamefont
  {Dimitrakopoulos}}, \bibinfo {author} {\bibfnamefont {K.~A.}\ \bibnamefont
  {Jenkins}}, \bibinfo {author} {\bibfnamefont {D.~B.}\ \bibnamefont {Farmer}},
  \bibinfo {author} {\bibfnamefont {H.-Y.}\ \bibnamefont {Chiu}}, \bibinfo
  {author} {\bibfnamefont {A.}~\bibnamefont {Grill}}, \ and\ \bibinfo {author}
  {\bibfnamefont {P.}~\bibnamefont {Avouris}},\ }\href {\doibase
  10.1126/science.327.5966.734-b} {\bibfield  {journal} {\bibinfo  {journal}
  {Science (New York, N.Y.)}\ }\textbf {\bibinfo {volume} {327}},\ \bibinfo
  {pages} {662} (\bibinfo {year} {2010})},\ \Eprint
  {http://arxiv.org/abs/1002.3845} {arXiv:1002.3845} \BibitemShut {NoStop}%
\bibitem [{\citenamefont {Morozov}\ \emph {et~al.}(2008)\citenamefont
  {Morozov}, \citenamefont {Novoselov}, \citenamefont {Katsnelson},
  \citenamefont {Schedin}, \citenamefont {Elias}, \citenamefont {Jaszczak},\
  and\ \citenamefont {Geim}}]{Morozov2008}%
  \BibitemOpen
  \bibfield  {author} {\bibinfo {author} {\bibfnamefont {S.~V.}\ \bibnamefont
  {Morozov}}, \bibinfo {author} {\bibfnamefont {K.~S.}\ \bibnamefont
  {Novoselov}}, \bibinfo {author} {\bibfnamefont {M.~I.}\ \bibnamefont
  {Katsnelson}}, \bibinfo {author} {\bibfnamefont {F.}~\bibnamefont {Schedin}},
  \bibinfo {author} {\bibfnamefont {D.~C.}\ \bibnamefont {Elias}}, \bibinfo
  {author} {\bibfnamefont {J.~A.}\ \bibnamefont {Jaszczak}}, \ and\ \bibinfo
  {author} {\bibfnamefont {A.~K.}\ \bibnamefont {Geim}},\ }\href {\doibase
  10.1103/PhysRevLett.100.016602} {\bibfield  {journal} {\bibinfo  {journal}
  {Physical Review Letters}\ }\textbf {\bibinfo {volume} {100}},\ \bibinfo
  {pages} {016602} (\bibinfo {year} {2008})},\ \Eprint
  {http://arxiv.org/abs/0710.5304} {arXiv:0710.5304} \BibitemShut {NoStop}%
\bibitem [{\citenamefont {Bolotin}\ \emph {et~al.}(2008)\citenamefont
  {Bolotin}, \citenamefont {Sikes}, \citenamefont {Jiang}, \citenamefont
  {Klima}, \citenamefont {Fudenberg}, \citenamefont {Hone}, \citenamefont
  {Kim},\ and\ \citenamefont {Stormer}}]{Bolotin2008}%
  \BibitemOpen
  \bibfield  {author} {\bibinfo {author} {\bibfnamefont {K.}~\bibnamefont
  {Bolotin}}, \bibinfo {author} {\bibfnamefont {K.}~\bibnamefont {Sikes}},
  \bibinfo {author} {\bibfnamefont {Z.}~\bibnamefont {Jiang}}, \bibinfo
  {author} {\bibfnamefont {M.}~\bibnamefont {Klima}}, \bibinfo {author}
  {\bibfnamefont {G.}~\bibnamefont {Fudenberg}}, \bibinfo {author}
  {\bibfnamefont {J.}~\bibnamefont {Hone}}, \bibinfo {author} {\bibfnamefont
  {P.}~\bibnamefont {Kim}}, \ and\ \bibinfo {author} {\bibfnamefont
  {H.}~\bibnamefont {Stormer}},\ }\href {\doibase 10.1016/j.ssc.2008.02.024}
  {\bibfield  {journal} {\bibinfo  {journal} {Solid State Communications}\
  }\textbf {\bibinfo {volume} {146}},\ \bibinfo {pages} {351} (\bibinfo {year}
  {2008})},\ \Eprint {http://arxiv.org/abs/0802.2389} {arXiv:0802.2389}
  \BibitemShut {NoStop}%
\bibitem [{\citenamefont {Berger}(2006)}]{Berger2006}%
  \BibitemOpen
  \bibfield  {author} {\bibinfo {author} {\bibfnamefont {C.}~\bibnamefont
  {Berger}},\ }\href {\doibase 10.1126/science.1125925} {\bibfield  {journal}
  {\bibinfo  {journal} {Science}\ }\textbf {\bibinfo {volume} {312}},\ \bibinfo
  {pages} {1191} (\bibinfo {year} {2006})}\BibitemShut {NoStop}%
\bibitem [{\citenamefont {Hibino}\ \emph {et~al.}(2008)\citenamefont {Hibino},
  \citenamefont {Kageshima}, \citenamefont {Maeda}, \citenamefont {Nagase},
  \citenamefont {Kobayashi},\ and\ \citenamefont {Yamaguchi}}]{Hibino2008}%
  \BibitemOpen
  \bibfield  {author} {\bibinfo {author} {\bibfnamefont {H.}~\bibnamefont
  {Hibino}}, \bibinfo {author} {\bibfnamefont {H.}~\bibnamefont {Kageshima}},
  \bibinfo {author} {\bibfnamefont {F.}~\bibnamefont {Maeda}}, \bibinfo
  {author} {\bibfnamefont {M.}~\bibnamefont {Nagase}}, \bibinfo {author}
  {\bibfnamefont {Y.}~\bibnamefont {Kobayashi}}, \ and\ \bibinfo {author}
  {\bibfnamefont {H.}~\bibnamefont {Yamaguchi}},\ }\href {\doibase
  10.1103/PhysRevB.77.075413} {\bibfield  {journal} {\bibinfo  {journal}
  {Physical Review B}\ }\textbf {\bibinfo {volume} {77}},\ \bibinfo {pages}
  {075413} (\bibinfo {year} {2008})},\ \Eprint {http://arxiv.org/abs/0710.0469}
  {arXiv:0710.0469} \BibitemShut {NoStop}%
\bibitem [{\citenamefont {Virojanadara}\ \emph {et~al.}(2008)\citenamefont
  {Virojanadara}, \citenamefont {Syv{\"{a}}jarvi}, \citenamefont {Yakimova},
  \citenamefont {Johansson}, \citenamefont {Zakharov},\ and\ \citenamefont
  {Balasubramanian}}]{Virojanadara2008}%
  \BibitemOpen
  \bibfield  {author} {\bibinfo {author} {\bibfnamefont {C.}~\bibnamefont
  {Virojanadara}}, \bibinfo {author} {\bibfnamefont {M.}~\bibnamefont
  {Syv{\"{a}}jarvi}}, \bibinfo {author} {\bibfnamefont {R.}~\bibnamefont
  {Yakimova}}, \bibinfo {author} {\bibfnamefont {L.~I.}\ \bibnamefont
  {Johansson}}, \bibinfo {author} {\bibfnamefont {A.~A.}\ \bibnamefont
  {Zakharov}}, \ and\ \bibinfo {author} {\bibfnamefont {T.}~\bibnamefont
  {Balasubramanian}},\ }\href {\doibase 10.1103/PhysRevB.78.245403} {\bibfield
  {journal} {\bibinfo  {journal} {Physical Review B}\ }\textbf {\bibinfo
  {volume} {78}},\ \bibinfo {pages} {245403} (\bibinfo {year}
  {2008})}\BibitemShut {NoStop}%
\bibitem [{\citenamefont {Norimatsu}\ and\ \citenamefont
  {Kusunoki}(2009)}]{Norimatsu2009}%
  \BibitemOpen
  \bibfield  {author} {\bibinfo {author} {\bibfnamefont {W.}~\bibnamefont
  {Norimatsu}}\ and\ \bibinfo {author} {\bibfnamefont {M.}~\bibnamefont
  {Kusunoki}},\ }\href {\doibase 10.1016/j.cplett.2008.11.095} {\bibfield
  {journal} {\bibinfo  {journal} {Chemical Physics Letters}\ }\textbf {\bibinfo
  {volume} {468}},\ \bibinfo {pages} {52} (\bibinfo {year} {2009})}\BibitemShut
  {NoStop}%
\bibitem [{\citenamefont {Emtsev}\ \emph {et~al.}(2009)\citenamefont {Emtsev},
  \citenamefont {Bostwick}, \citenamefont {Horn}, \citenamefont {Jobst},
  \citenamefont {Kellogg}, \citenamefont {Ley}, \citenamefont {McChesney},
  \citenamefont {Ohta}, \citenamefont {Reshanov}, \citenamefont {R{\"{o}}hrl},
  \citenamefont {Rotenberg}, \citenamefont {Schmid}, \citenamefont {Waldmann},
  \citenamefont {Weber},\ and\ \citenamefont {Seyller}}]{Emtsev2009}%
  \BibitemOpen
  \bibfield  {author} {\bibinfo {author} {\bibfnamefont {K.~V.}\ \bibnamefont
  {Emtsev}}, \bibinfo {author} {\bibfnamefont {A.}~\bibnamefont {Bostwick}},
  \bibinfo {author} {\bibfnamefont {K.}~\bibnamefont {Horn}}, \bibinfo {author}
  {\bibfnamefont {J.}~\bibnamefont {Jobst}}, \bibinfo {author} {\bibfnamefont
  {G.~L.}\ \bibnamefont {Kellogg}}, \bibinfo {author} {\bibfnamefont
  {L.}~\bibnamefont {Ley}}, \bibinfo {author} {\bibfnamefont {J.~L.}\
  \bibnamefont {McChesney}}, \bibinfo {author} {\bibfnamefont {T.}~\bibnamefont
  {Ohta}}, \bibinfo {author} {\bibfnamefont {S.~a.}\ \bibnamefont {Reshanov}},
  \bibinfo {author} {\bibfnamefont {J.}~\bibnamefont {R{\"{o}}hrl}}, \bibinfo
  {author} {\bibfnamefont {E.}~\bibnamefont {Rotenberg}}, \bibinfo {author}
  {\bibfnamefont {A.~K.}\ \bibnamefont {Schmid}}, \bibinfo {author}
  {\bibfnamefont {D.}~\bibnamefont {Waldmann}}, \bibinfo {author}
  {\bibfnamefont {H.~B.}\ \bibnamefont {Weber}}, \ and\ \bibinfo {author}
  {\bibfnamefont {T.}~\bibnamefont {Seyller}},\ }\href {\doibase
  10.1038/nmat2382} {\bibfield  {journal} {\bibinfo  {journal} {Nature
  Materials}\ }\textbf {\bibinfo {volume} {8}},\ \bibinfo {pages} {203}
  (\bibinfo {year} {2009})},\ \Eprint {http://arxiv.org/abs/arXiv:0903.2067v1}
  {arXiv:arXiv:0903.2067v1} \BibitemShut {NoStop}%
\bibitem [{\citenamefont {Hibino}\ \emph {et~al.}(2010)\citenamefont {Hibino},
  \citenamefont {Kageshima},\ and\ \citenamefont
  {Nagase}}]{HibinoH.KageshimaH.&Nagase2010}%
  \BibitemOpen
  \bibfield  {author} {\bibinfo {author} {\bibfnamefont {H.}~\bibnamefont
  {Hibino}}, \bibinfo {author} {\bibfnamefont {H.}~\bibnamefont {Kageshima}}, \
  and\ \bibinfo {author} {\bibfnamefont {M.}~\bibnamefont {Nagase}},\ }\href
  {\doibase 10.1088/0022-3727/43/37/374005} {\bibfield  {journal} {\bibinfo
  {journal} {Journal of Physics D: Applied Physics}\ }\textbf {\bibinfo
  {volume} {43}},\ \bibinfo {pages} {374005} (\bibinfo {year}
  {2010})}\BibitemShut {NoStop}%
\bibitem [{\citenamefont {Hannon}\ and\ \citenamefont
  {Tromp}(2008)}]{Hannon2008}%
  \BibitemOpen
  \bibfield  {author} {\bibinfo {author} {\bibfnamefont {J.~B.}\ \bibnamefont
  {Hannon}}\ and\ \bibinfo {author} {\bibfnamefont {R.~M.}\ \bibnamefont
  {Tromp}},\ }\href {\doibase 10.1103/PhysRevB.77.241404} {\bibfield  {journal}
  {\bibinfo  {journal} {Physical Review B}\ }\textbf {\bibinfo {volume} {77}},\
  \bibinfo {pages} {241404} (\bibinfo {year} {2008})}\BibitemShut {NoStop}%
\bibitem [{\citenamefont {Norimatsu}\ and\ \citenamefont
  {Kusunoki}(2010)}]{Norimatsu2010}%
  \BibitemOpen
  \bibfield  {author} {\bibinfo {author} {\bibfnamefont {W.}~\bibnamefont
  {Norimatsu}}\ and\ \bibinfo {author} {\bibfnamefont {M.}~\bibnamefont
  {Kusunoki}},\ }\href {\doibase 10.1016/j.physe.2009.11.151} {\bibfield
  {journal} {\bibinfo  {journal} {Physica E: Low-dimensional Systems and
  Nanostructures}\ }\textbf {\bibinfo {volume} {42}},\ \bibinfo {pages} {691}
  (\bibinfo {year} {2010})}\BibitemShut {NoStop}%
\bibitem [{\citenamefont {Norimatsu}\ and\ \citenamefont
  {Kusunoki}(2014{\natexlab{a}})}]{Norimatsu2014a}%
  \BibitemOpen
  \bibfield  {author} {\bibinfo {author} {\bibfnamefont {W.}~\bibnamefont
  {Norimatsu}}\ and\ \bibinfo {author} {\bibfnamefont {M.}~\bibnamefont
  {Kusunoki}},\ }\href {\doibase 10.1088/0022-3727/47/9/094017} {\bibfield
  {journal} {\bibinfo  {journal} {Journal of Physics D: Applied Physics}\
  }\textbf {\bibinfo {volume} {47}},\ \bibinfo {pages} {094017} (\bibinfo
  {year} {2014}{\natexlab{a}})}\BibitemShut {NoStop}%
\bibitem [{\citenamefont {Chen}\ \emph {et~al.}(2005)\citenamefont {Chen},
  \citenamefont {Xu}, \citenamefont {Liu}, \citenamefont {Gao}, \citenamefont
  {Qi}, \citenamefont {Peng}, \citenamefont {Tan}, \citenamefont {Feng},
  \citenamefont {Loh},\ and\ \citenamefont {Wee}}]{Chen2005}%
  \BibitemOpen
  \bibfield  {author} {\bibinfo {author} {\bibfnamefont {W.}~\bibnamefont
  {Chen}}, \bibinfo {author} {\bibfnamefont {H.}~\bibnamefont {Xu}}, \bibinfo
  {author} {\bibfnamefont {L.}~\bibnamefont {Liu}}, \bibinfo {author}
  {\bibfnamefont {X.}~\bibnamefont {Gao}}, \bibinfo {author} {\bibfnamefont
  {D.}~\bibnamefont {Qi}}, \bibinfo {author} {\bibfnamefont {G.}~\bibnamefont
  {Peng}}, \bibinfo {author} {\bibfnamefont {S.~C.}\ \bibnamefont {Tan}},
  \bibinfo {author} {\bibfnamefont {Y.}~\bibnamefont {Feng}}, \bibinfo {author}
  {\bibfnamefont {K.~P.}\ \bibnamefont {Loh}}, \ and\ \bibinfo {author}
  {\bibfnamefont {A.~T.~S.}\ \bibnamefont {Wee}},\ }\href {\doibase
  10.1016/j.susc.2005.09.013} {\bibfield  {journal} {\bibinfo  {journal}
  {Surface Science}\ }\textbf {\bibinfo {volume} {596}},\ \bibinfo {pages}
  {176} (\bibinfo {year} {2005})}\BibitemShut {NoStop}%
\bibitem [{\citenamefont {Tsai}\ \emph {et~al.}(1992)\citenamefont {Tsai},
  \citenamefont {Chang}, \citenamefont {Dow},\ and\ \citenamefont
  {Tsong}}]{Tsai1992}%
  \BibitemOpen
  \bibfield  {author} {\bibinfo {author} {\bibfnamefont {M.-H.}\ \bibnamefont
  {Tsai}}, \bibinfo {author} {\bibfnamefont {C.~S.}\ \bibnamefont {Chang}},
  \bibinfo {author} {\bibfnamefont {J.~D.}\ \bibnamefont {Dow}}, \ and\
  \bibinfo {author} {\bibfnamefont {I.~S.~T.}\ \bibnamefont {Tsong}},\ }\href
  {\doibase 10.1103/PhysRevB.45.1327} {\bibfield  {journal} {\bibinfo
  {journal} {Physical Review B}\ }\textbf {\bibinfo {volume} {45}},\ \bibinfo
  {pages} {1327} (\bibinfo {year} {1992})}\BibitemShut {NoStop}%
\bibitem [{\citenamefont {Seyller}\ \emph {et~al.}(2006)\citenamefont
  {Seyller}, \citenamefont {Emtsev}, \citenamefont {Gao}, \citenamefont
  {Speck}, \citenamefont {Ley}, \citenamefont {Tadich}, \citenamefont
  {Broekman}, \citenamefont {Riley}, \citenamefont {Leckey}, \citenamefont
  {Rader}, \citenamefont {Varykhalov},\ and\ \citenamefont
  {Shikin}}]{Seyller2006}%
  \BibitemOpen
  \bibfield  {author} {\bibinfo {author} {\bibfnamefont {T.}~\bibnamefont
  {Seyller}}, \bibinfo {author} {\bibfnamefont {K.}~\bibnamefont {Emtsev}},
  \bibinfo {author} {\bibfnamefont {K.}~\bibnamefont {Gao}}, \bibinfo {author}
  {\bibfnamefont {F.}~\bibnamefont {Speck}}, \bibinfo {author} {\bibfnamefont
  {L.}~\bibnamefont {Ley}}, \bibinfo {author} {\bibfnamefont {A.}~\bibnamefont
  {Tadich}}, \bibinfo {author} {\bibfnamefont {L.}~\bibnamefont {Broekman}},
  \bibinfo {author} {\bibfnamefont {J.}~\bibnamefont {Riley}}, \bibinfo
  {author} {\bibfnamefont {R.}~\bibnamefont {Leckey}}, \bibinfo {author}
  {\bibfnamefont {O.}~\bibnamefont {Rader}}, \bibinfo {author} {\bibfnamefont
  {A.}~\bibnamefont {Varykhalov}}, \ and\ \bibinfo {author} {\bibfnamefont
  {A.}~\bibnamefont {Shikin}},\ }\href {\doibase 10.1016/j.susc.2006.01.102}
  {\bibfield  {journal} {\bibinfo  {journal} {Surface Science}\ }\textbf
  {\bibinfo {volume} {600}},\ \bibinfo {pages} {3906} (\bibinfo {year}
  {2006})}\BibitemShut {NoStop}%
\bibitem [{\citenamefont {Charrier}\ \emph {et~al.}(2002)\citenamefont
  {Charrier}, \citenamefont {Coati}, \citenamefont {Argunova}, \citenamefont
  {Thibaudau}, \citenamefont {Garreau}, \citenamefont {Pinchaux}, \citenamefont
  {Forbeaux}, \citenamefont {Debever}, \citenamefont {Sauvage-Simkin},\ and\
  \citenamefont {Themlin}}]{Charrier2002}%
  \BibitemOpen
  \bibfield  {author} {\bibinfo {author} {\bibfnamefont {A.}~\bibnamefont
  {Charrier}}, \bibinfo {author} {\bibfnamefont {A.}~\bibnamefont {Coati}},
  \bibinfo {author} {\bibfnamefont {T.}~\bibnamefont {Argunova}}, \bibinfo
  {author} {\bibfnamefont {F.}~\bibnamefont {Thibaudau}}, \bibinfo {author}
  {\bibfnamefont {Y.}~\bibnamefont {Garreau}}, \bibinfo {author} {\bibfnamefont
  {R.}~\bibnamefont {Pinchaux}}, \bibinfo {author} {\bibfnamefont
  {I.}~\bibnamefont {Forbeaux}}, \bibinfo {author} {\bibfnamefont {J.-M.}\
  \bibnamefont {Debever}}, \bibinfo {author} {\bibfnamefont {M.}~\bibnamefont
  {Sauvage-Simkin}}, \ and\ \bibinfo {author} {\bibfnamefont {J.-M.}\
  \bibnamefont {Themlin}},\ }\href {\doibase 10.1063/1.1498962} {\bibfield
  {journal} {\bibinfo  {journal} {Journal of Applied Physics}\ }\textbf
  {\bibinfo {volume} {92}},\ \bibinfo {pages} {2479} (\bibinfo {year}
  {2002})}\BibitemShut {NoStop}%
\bibitem [{\citenamefont {Rollings}\ \emph {et~al.}(2006)\citenamefont
  {Rollings}, \citenamefont {Gweon}, \citenamefont {Zhou}, \citenamefont {Mun},
  \citenamefont {McChesney}, \citenamefont {Hussain}, \citenamefont {Fedorov},
  \citenamefont {First}, \citenamefont {de~Heer},\ and\ \citenamefont
  {Lanzara}}]{Rollings2006}%
  \BibitemOpen
  \bibfield  {author} {\bibinfo {author} {\bibfnamefont {E.}~\bibnamefont
  {Rollings}}, \bibinfo {author} {\bibfnamefont {G.-H.}\ \bibnamefont {Gweon}},
  \bibinfo {author} {\bibfnamefont {S.}~\bibnamefont {Zhou}}, \bibinfo {author}
  {\bibfnamefont {B.}~\bibnamefont {Mun}}, \bibinfo {author} {\bibfnamefont
  {J.}~\bibnamefont {McChesney}}, \bibinfo {author} {\bibfnamefont
  {B.}~\bibnamefont {Hussain}}, \bibinfo {author} {\bibfnamefont
  {A.}~\bibnamefont {Fedorov}}, \bibinfo {author} {\bibfnamefont
  {P.}~\bibnamefont {First}}, \bibinfo {author} {\bibfnamefont
  {W.}~\bibnamefont {de~Heer}}, \ and\ \bibinfo {author} {\bibfnamefont
  {A.}~\bibnamefont {Lanzara}},\ }\href {\doibase 10.1016/j.jpcs.2006.05.010}
  {\bibfield  {journal} {\bibinfo  {journal} {Journal of Physics and Chemistry
  of Solids}\ }\textbf {\bibinfo {volume} {67}},\ \bibinfo {pages} {2172}
  (\bibinfo {year} {2006})},\ \Eprint {http://arxiv.org/abs/0512226}
  {arXiv:0512226 [cond-mat]} \BibitemShut {NoStop}%
\bibitem [{\citenamefont {Varchon}\ \emph {et~al.}(2008)\citenamefont
  {Varchon}, \citenamefont {Mallet}, \citenamefont {Veuillen},\ and\
  \citenamefont {Magaud}}]{Varchon2008}%
  \BibitemOpen
  \bibfield  {author} {\bibinfo {author} {\bibfnamefont {F.}~\bibnamefont
  {Varchon}}, \bibinfo {author} {\bibfnamefont {P.}~\bibnamefont {Mallet}},
  \bibinfo {author} {\bibfnamefont {J.-Y.}\ \bibnamefont {Veuillen}}, \ and\
  \bibinfo {author} {\bibfnamefont {L.}~\bibnamefont {Magaud}},\ }\href
  {\doibase 10.1103/PhysRevB.77.235412} {\bibfield  {journal} {\bibinfo
  {journal} {Physical Review B}\ }\textbf {\bibinfo {volume} {77}},\ \bibinfo
  {pages} {235412} (\bibinfo {year} {2008})}\BibitemShut {NoStop}%
\bibitem [{\citenamefont {Tetlow}\ \emph {et~al.}(2014)\citenamefont {Tetlow},
  \citenamefont {{Posthuma de Boer}}, \citenamefont {Ford}, \citenamefont
  {Vvedensky}, \citenamefont {Coraux},\ and\ \citenamefont
  {Kantorovich}}]{Tetlow2014}%
  \BibitemOpen
  \bibfield  {author} {\bibinfo {author} {\bibfnamefont {H.}~\bibnamefont
  {Tetlow}}, \bibinfo {author} {\bibfnamefont {J.}~\bibnamefont {{Posthuma de
  Boer}}}, \bibinfo {author} {\bibfnamefont {I.~J.}\ \bibnamefont {Ford}},
  \bibinfo {author} {\bibfnamefont {D.~D.}\ \bibnamefont {Vvedensky}}, \bibinfo
  {author} {\bibfnamefont {J.}~\bibnamefont {Coraux}}, \ and\ \bibinfo {author}
  {\bibfnamefont {L.}~\bibnamefont {Kantorovich}},\ }\href {\doibase
  10.1016/j.physrep.2014.03.003} {\bibfield  {journal} {\bibinfo  {journal}
  {Physics Reports}\ }\textbf {\bibinfo {volume} {542}},\ \bibinfo {pages}
  {195} (\bibinfo {year} {2014})},\ \Eprint {http://arxiv.org/abs/1602.06707}
  {arXiv:1602.06707} \BibitemShut {NoStop}%
\bibitem [{\citenamefont {Norimatsu}\ and\ \citenamefont
  {Kusunoki}(2014{\natexlab{b}})}]{Norimatsu2014}%
  \BibitemOpen
  \bibfield  {author} {\bibinfo {author} {\bibfnamefont {W.}~\bibnamefont
  {Norimatsu}}\ and\ \bibinfo {author} {\bibfnamefont {M.}~\bibnamefont
  {Kusunoki}},\ }\href {\doibase 10.1088/0268-1242/29/6/064009} {\bibfield
  {journal} {\bibinfo  {journal} {Semiconductor Science and Technology}\
  }\textbf {\bibinfo {volume} {29}},\ \bibinfo {pages} {064009} (\bibinfo
  {year} {2014}{\natexlab{b}})}\BibitemShut {NoStop}%
\bibitem [{\citenamefont {Kageshima}\ \emph {et~al.}(2011)\citenamefont
  {Kageshima}, \citenamefont {Hibino}, \citenamefont {Yamaguchi},\ and\
  \citenamefont {Nagase}}]{Kageshima2011}%
  \BibitemOpen
  \bibfield  {author} {\bibinfo {author} {\bibfnamefont {H.}~\bibnamefont
  {Kageshima}}, \bibinfo {author} {\bibfnamefont {H.}~\bibnamefont {Hibino}},
  \bibinfo {author} {\bibfnamefont {H.}~\bibnamefont {Yamaguchi}}, \ and\
  \bibinfo {author} {\bibfnamefont {M.}~\bibnamefont {Nagase}},\ }\href
  {\doibase 10.1143/JJAP.50.095601} {\bibfield  {journal} {\bibinfo  {journal}
  {Japanese Journal of Applied Physics}\ }\textbf {\bibinfo {volume} {50}},\
  \bibinfo {pages} {095601} (\bibinfo {year} {2011})}\BibitemShut {NoStop}%
\bibitem [{\citenamefont {Kageshima}\ \emph {et~al.}(2012)\citenamefont
  {Kageshima}, \citenamefont {Hibino},\ and\ \citenamefont
  {Tanabe}}]{Kageshima2012}%
  \BibitemOpen
  \bibfield  {author} {\bibinfo {author} {\bibfnamefont {H.}~\bibnamefont
  {Kageshima}}, \bibinfo {author} {\bibfnamefont {H.}~\bibnamefont {Hibino}}, \
  and\ \bibinfo {author} {\bibfnamefont {S.}~\bibnamefont {Tanabe}},\ }\href
  {\doibase 10.1088/0953-8984/24/31/314215} {\bibfield  {journal} {\bibinfo
  {journal} {Journal of Physics: Condensed Matter}\ }\textbf {\bibinfo {volume}
  {24}},\ \bibinfo {pages} {314215} (\bibinfo {year} {2012})}\BibitemShut
  {NoStop}%
\bibitem [{\citenamefont {Inoue}\ \emph {et~al.}(2012)\citenamefont {Inoue},
  \citenamefont {Kageshima}, \citenamefont {Kangawa},\ and\ \citenamefont
  {Kakimoto}}]{Inoue2012}%
  \BibitemOpen
  \bibfield  {author} {\bibinfo {author} {\bibfnamefont {M.}~\bibnamefont
  {Inoue}}, \bibinfo {author} {\bibfnamefont {H.}~\bibnamefont {Kageshima}},
  \bibinfo {author} {\bibfnamefont {Y.}~\bibnamefont {Kangawa}}, \ and\
  \bibinfo {author} {\bibfnamefont {K.}~\bibnamefont {Kakimoto}},\ }\href
  {\doibase 10.1103/PhysRevB.86.085417} {\bibfield  {journal} {\bibinfo
  {journal} {Physical Review B}\ }\textbf {\bibinfo {volume} {86}},\ \bibinfo
  {pages} {085417} (\bibinfo {year} {2012})}\BibitemShut {NoStop}%
\bibitem [{\citenamefont {Morita}\ \emph {et~al.}(2013)\citenamefont {Morita},
  \citenamefont {Norimatsu}, \citenamefont {Qian}, \citenamefont {Irle},\ and\
  \citenamefont {Kusunoki}}]{Morita2013}%
  \BibitemOpen
  \bibfield  {author} {\bibinfo {author} {\bibfnamefont {M.}~\bibnamefont
  {Morita}}, \bibinfo {author} {\bibfnamefont {W.}~\bibnamefont {Norimatsu}},
  \bibinfo {author} {\bibfnamefont {H.-J.}\ \bibnamefont {Qian}}, \bibinfo
  {author} {\bibfnamefont {S.}~\bibnamefont {Irle}}, \ and\ \bibinfo {author}
  {\bibfnamefont {M.}~\bibnamefont {Kusunoki}},\ }\href {\doibase
  10.1063/1.4824425} {\bibfield  {journal} {\bibinfo  {journal} {Applied
  Physics Letters}\ }\textbf {\bibinfo {volume} {103}},\ \bibinfo {pages}
  {141602} (\bibinfo {year} {2013})}\BibitemShut {NoStop}%
\bibitem [{\citenamefont {Lampin}\ \emph {et~al.}(2010)\citenamefont {Lampin},
  \citenamefont {Priester}, \citenamefont {Krzeminski},\ and\ \citenamefont
  {Magaud}}]{Lampin2010}%
  \BibitemOpen
  \bibfield  {author} {\bibinfo {author} {\bibfnamefont {E.}~\bibnamefont
  {Lampin}}, \bibinfo {author} {\bibfnamefont {C.}~\bibnamefont {Priester}},
  \bibinfo {author} {\bibfnamefont {C.}~\bibnamefont {Krzeminski}}, \ and\
  \bibinfo {author} {\bibfnamefont {L.}~\bibnamefont {Magaud}},\ }\href
  {\doibase 10.1063/1.3357297} {\bibfield  {journal} {\bibinfo  {journal}
  {Journal of Applied Physics}\ }\textbf {\bibinfo {volume} {107}},\ \bibinfo
  {pages} {103514} (\bibinfo {year} {2010})},\ \Eprint
  {http://arxiv.org/abs/0912.2034} {arXiv:0912.2034} \BibitemShut {NoStop}%
\bibitem [{\citenamefont {Tang}\ \emph
  {et~al.}(2008{\natexlab{a}})\citenamefont {Tang}, \citenamefont {Meng},
  \citenamefont {Sun}, \citenamefont {Zhang},\ and\ \citenamefont
  {Zhong}}]{Tang2008a}%
  \BibitemOpen
  \bibfield  {author} {\bibinfo {author} {\bibfnamefont {C.}~\bibnamefont
  {Tang}}, \bibinfo {author} {\bibfnamefont {L.}~\bibnamefont {Meng}}, \bibinfo
  {author} {\bibfnamefont {L.}~\bibnamefont {Sun}}, \bibinfo {author}
  {\bibfnamefont {K.}~\bibnamefont {Zhang}}, \ and\ \bibinfo {author}
  {\bibfnamefont {J.}~\bibnamefont {Zhong}},\ }\href {\doibase
  10.1063/1.3032895} {\bibfield  {journal} {\bibinfo  {journal} {Journal of
  Applied Physics}\ }\textbf {\bibinfo {volume} {104}},\ \bibinfo {pages}
  {113536} (\bibinfo {year} {2008}{\natexlab{a}})}\BibitemShut {NoStop}%
\bibitem [{\citenamefont {Tang}\ \emph
  {et~al.}(2008{\natexlab{b}})\citenamefont {Tang}, \citenamefont {Meng},
  \citenamefont {Xiao},\ and\ \citenamefont {Zhong}}]{Tang2008}%
  \BibitemOpen
  \bibfield  {author} {\bibinfo {author} {\bibfnamefont {C.}~\bibnamefont
  {Tang}}, \bibinfo {author} {\bibfnamefont {L.}~\bibnamefont {Meng}}, \bibinfo
  {author} {\bibfnamefont {H.}~\bibnamefont {Xiao}}, \ and\ \bibinfo {author}
  {\bibfnamefont {J.}~\bibnamefont {Zhong}},\ }\href {\doibase
  10.1063/1.2894728} {\bibfield  {journal} {\bibinfo  {journal} {Journal of
  Applied Physics}\ }\textbf {\bibinfo {volume} {103}} (\bibinfo {year}
  {2008}{\natexlab{b}}),\ 10.1063/1.2894728}\BibitemShut {NoStop}%
\bibitem [{\citenamefont {Jakse}\ \emph {et~al.}(2011)\citenamefont {Jakse},
  \citenamefont {Arifin},\ and\ \citenamefont {Lai}}]{Jakse2011}%
  \BibitemOpen
  \bibfield  {author} {\bibinfo {author} {\bibnamefont {Jakse}}, \bibinfo
  {author} {\bibnamefont {Arifin}}, \ and\ \bibinfo {author} {\bibnamefont
  {Lai}},\ }\href {\doibase 10.5488/CMP.14.43802} {\bibfield  {journal}
  {\bibinfo  {journal} {Condensed Matter Physics}\ }\textbf {\bibinfo {volume}
  {14}},\ \bibinfo {pages} {43802} (\bibinfo {year} {2011})},\ \Eprint
  {http://arxiv.org/abs/arXiv:1202.4846v1} {arXiv:arXiv:1202.4846v1}
  \BibitemShut {NoStop}%
\bibitem [{\citenamefont {Ono}\ \emph {et~al.}(2016)\citenamefont {Ono},
  \citenamefont {Yamasaki}, \citenamefont {Nara},\ and\ \citenamefont
  {Ohno}}]{Ono2016graphene}%
  \BibitemOpen
  \bibfield  {author} {\bibinfo {author} {\bibfnamefont {Y.}~\bibnamefont
  {Ono}}, \bibinfo {author} {\bibfnamefont {T.}~\bibnamefont {Yamasaki}},
  \bibinfo {author} {\bibfnamefont {J.}~\bibnamefont {Nara}}, \ and\ \bibinfo
  {author} {\bibfnamefont {T.}~\bibnamefont {Ohno}},\ }in\ \href@noop {} {\emph
  {\bibinfo {booktitle} {NT16 The 17th International Conference on the Science
  and Application of Nanotubes}}}\ (\bibinfo {year} {2016})\ pp.\ \bibinfo
  {pages} {165--166}\BibitemShut {NoStop}%
\bibitem [{\citenamefont {Takamoto}\ \emph {et~al.}(2016)\citenamefont
  {Takamoto}, \citenamefont {Kumagai}, \citenamefont {Yamasaki}, \citenamefont
  {Ohno}, \citenamefont {Kaneta}, \citenamefont {Hatano},\ and\ \citenamefont
  {Izumi}}]{Takamoto2016}%
  \BibitemOpen
  \bibfield  {author} {\bibinfo {author} {\bibfnamefont {S.}~\bibnamefont
  {Takamoto}}, \bibinfo {author} {\bibfnamefont {T.}~\bibnamefont {Kumagai}},
  \bibinfo {author} {\bibfnamefont {T.}~\bibnamefont {Yamasaki}}, \bibinfo
  {author} {\bibfnamefont {T.}~\bibnamefont {Ohno}}, \bibinfo {author}
  {\bibfnamefont {C.}~\bibnamefont {Kaneta}}, \bibinfo {author} {\bibfnamefont
  {A.}~\bibnamefont {Hatano}}, \ and\ \bibinfo {author} {\bibfnamefont
  {S.}~\bibnamefont {Izumi}},\ }\href {\doibase 10.1063/1.4965863} {\bibfield
  {journal} {\bibinfo  {journal} {Journal of Applied Physics}\ }\textbf
  {\bibinfo {volume} {120}},\ \bibinfo {pages} {165109} (\bibinfo {year}
  {2016})}\BibitemShut {NoStop}%
\bibitem [{\citenamefont {Kumagai}\ \emph {et~al.}(2004)\citenamefont
  {Kumagai}, \citenamefont {Izumi}, \citenamefont {Hara},\ and\ \citenamefont
  {Sakai}}]{KumagaiSiO2}%
  \BibitemOpen
  \bibfield  {author} {\bibinfo {author} {\bibfnamefont {T.}~\bibnamefont
  {Kumagai}}, \bibinfo {author} {\bibfnamefont {S.}~\bibnamefont {Izumi}},
  \bibinfo {author} {\bibfnamefont {S.}~\bibnamefont {Hara}}, \ and\ \bibinfo
  {author} {\bibfnamefont {S.}~\bibnamefont {Sakai}},\ }in\ \href@noop {}
  {\emph {\bibinfo {booktitle} {Extended Abstracts of International Conference
  on Computational Methods}}}\ (\bibinfo {year} {2004})\ p.~\bibinfo {pages}
  {75}\BibitemShut {NoStop}%
\bibitem [{\citenamefont {Yu}\ \emph {et~al.}(2007)\citenamefont {Yu},
  \citenamefont {Sinnott},\ and\ \citenamefont {Phillpot}}]{Yu2007}%
  \BibitemOpen
  \bibfield  {author} {\bibinfo {author} {\bibfnamefont {J.}~\bibnamefont
  {Yu}}, \bibinfo {author} {\bibfnamefont {S.~B.}\ \bibnamefont {Sinnott}}, \
  and\ \bibinfo {author} {\bibfnamefont {S.~R.}\ \bibnamefont {Phillpot}},\
  }\href {\doibase 10.1103/PhysRevB.75.085311} {\bibfield  {journal} {\bibinfo
  {journal} {Physical Review B}\ }\textbf {\bibinfo {volume} {75}},\ \bibinfo
  {pages} {085311} (\bibinfo {year} {2007})}\BibitemShut {NoStop}%
\bibitem [{\citenamefont {van Duin}\ \emph {et~al.}(2003)\citenamefont {van
  Duin}, \citenamefont {Strachan}, \citenamefont {Stewman}, \citenamefont
  {Zhang}, \citenamefont {Xu},\ and\ \citenamefont {Goddard}}]{VanDuin2003}%
  \BibitemOpen
  \bibfield  {author} {\bibinfo {author} {\bibfnamefont {A.~C.~T.}\
  \bibnamefont {van Duin}}, \bibinfo {author} {\bibfnamefont {A.}~\bibnamefont
  {Strachan}}, \bibinfo {author} {\bibfnamefont {S.}~\bibnamefont {Stewman}},
  \bibinfo {author} {\bibfnamefont {Q.}~\bibnamefont {Zhang}}, \bibinfo
  {author} {\bibfnamefont {X.}~\bibnamefont {Xu}}, \ and\ \bibinfo {author}
  {\bibfnamefont {W.~A.}\ \bibnamefont {Goddard}},\ }\href {\doibase
  10.1021/jp0276303} {\bibfield  {journal} {\bibinfo  {journal} {The Journal of
  Physical Chemistry A}\ }\textbf {\bibinfo {volume} {107}},\ \bibinfo {pages}
  {3803} (\bibinfo {year} {2003})}\BibitemShut {NoStop}%
\bibitem [{\citenamefont {Tersoff}(1986)}]{Tersoff1986}%
  \BibitemOpen
  \bibfield  {author} {\bibinfo {author} {\bibfnamefont {J.}~\bibnamefont
  {Tersoff}},\ }\href {\doibase 10.1103/PhysRevLett.56.632} {\bibfield
  {journal} {\bibinfo  {journal} {Physical Review Letters}\ }\textbf {\bibinfo
  {volume} {56}},\ \bibinfo {pages} {632} (\bibinfo {year} {1986})}\BibitemShut
  {NoStop}%
\bibitem [{\citenamefont {Tersoff}(1988{\natexlab{a}})}]{Tersoff1988a}%
  \BibitemOpen
  \bibfield  {author} {\bibinfo {author} {\bibfnamefont {J.}~\bibnamefont
  {Tersoff}},\ }\href {\doibase 10.1103/PhysRevB.37.6991} {\bibfield  {journal}
  {\bibinfo  {journal} {Physical Review B}\ }\textbf {\bibinfo {volume} {37}},\
  \bibinfo {pages} {6991} (\bibinfo {year} {1988}{\natexlab{a}})}\BibitemShut
  {NoStop}%
\bibitem [{\citenamefont {Tersoff}(1988{\natexlab{b}})}]{Tersoff1988b}%
  \BibitemOpen
  \bibfield  {author} {\bibinfo {author} {\bibfnamefont {J.}~\bibnamefont
  {Tersoff}},\ }\href {\doibase 10.1103/PhysRevLett.61.2879} {\bibfield
  {journal} {\bibinfo  {journal} {Physical Review Letters}\ }\textbf {\bibinfo
  {volume} {61}},\ \bibinfo {pages} {2879} (\bibinfo {year}
  {1988}{\natexlab{b}})}\BibitemShut {NoStop}%
\bibitem [{\citenamefont {Plimpton}(1995)}]{plimpton1995fast}%
  \BibitemOpen
  \bibfield  {author} {\bibinfo {author} {\bibfnamefont {S.}~\bibnamefont
  {Plimpton}},\ }\href {\doibase https://doi.org/10.1006/jcph.1995.1039"}
  {\bibfield  {journal} {\bibinfo  {journal} {Journal of Computational
  Physics}\ }\textbf {\bibinfo {volume} {117}},\ \bibinfo {pages} {1} (\bibinfo
  {year} {1995})}\BibitemShut {NoStop}%
\bibitem [{lam()}]{lammpssource}%
  \BibitemOpen
  \href@noop {} {}\bibinfo {note} {The implementation and parameters are
  available at \url{https://github.com/Takamoto-So/tersoff_k2}.}\BibitemShut
  {Stop}%
\bibitem [{\citenamefont {Kumagai}\ and\ \citenamefont
  {Izumi}(2011)}]{KumagaiFitting}%
  \BibitemOpen
  \bibfield  {author} {\bibinfo {author} {\bibfnamefont {T.}~\bibnamefont
  {Kumagai}}\ and\ \bibinfo {author} {\bibfnamefont {S.}~\bibnamefont
  {Izumi}},\ }\href@noop {} {\bibfield  {journal} {\bibinfo  {journal}
  {Transactions of the Japan Society of Mechanical Engineers}\ }\bibinfo
  {series} {A},\ \textbf {\bibinfo {volume} {77}},\ \bibinfo {pages} {2026}
  (\bibinfo {year} {2011})},\ \bibinfo {note} {(in Japanese)}\BibitemShut
  {NoStop}%
\bibitem [{nim()}]{nimsazuma}%
  \BibitemOpen
  \href@noop {} {\enquote {\bibinfo {title} {{NIMS} / nano-simulation
  software},}\ }\bibinfo {howpublished}
  {\url{https://azuma.nims.go.jp}}\BibitemShut {NoStop}%
\bibitem [{\citenamefont {Ohno}\ \emph {et~al.}(2007)\citenamefont {Ohno},
  \citenamefont {Yamamoto}, \citenamefont {Kokubo}, \citenamefont {Azami},
  \citenamefont {Sakaguchi}, \citenamefont {Uda}, \citenamefont {Yamasaki},
  \citenamefont {Fukata},\ and\ \citenamefont
  {Koga}}]{Ohno:2007:FCL:1362622.1362699}%
  \BibitemOpen
  \bibfield  {author} {\bibinfo {author} {\bibfnamefont {T.}~\bibnamefont
  {Ohno}}, \bibinfo {author} {\bibfnamefont {T.}~\bibnamefont {Yamamoto}},
  \bibinfo {author} {\bibfnamefont {T.}~\bibnamefont {Kokubo}}, \bibinfo
  {author} {\bibfnamefont {A.}~\bibnamefont {Azami}}, \bibinfo {author}
  {\bibfnamefont {Y.}~\bibnamefont {Sakaguchi}}, \bibinfo {author}
  {\bibfnamefont {T.}~\bibnamefont {Uda}}, \bibinfo {author} {\bibfnamefont
  {T.}~\bibnamefont {Yamasaki}}, \bibinfo {author} {\bibfnamefont
  {D.}~\bibnamefont {Fukata}}, \ and\ \bibinfo {author} {\bibfnamefont
  {J.}~\bibnamefont {Koga}},\ }in\ \href {\doibase 10.1145/1362622.1362699}
  {\emph {\bibinfo {booktitle} {Proceedings of the 2007 ACM/IEEE Conference on
  Supercomputing}}},\ \bibinfo {series and number} {SC '07}\ (\bibinfo
  {publisher} {ACM},\ \bibinfo {address} {New York, NY, USA},\ \bibinfo {year}
  {2007})\ pp.\ \bibinfo {pages} {57:1--57:6}\BibitemShut {NoStop}%
\bibitem [{\citenamefont {Tersoff}(1994)}]{J.Tersoff1994}%
  \BibitemOpen
  \bibfield  {author} {\bibinfo {author} {\bibfnamefont {J.}~\bibnamefont
  {Tersoff}},\ }\href {\doibase 10.1103/PhysRevB.49.16349} {\bibfield
  {journal} {\bibinfo  {journal} {Physical Review B}\ }\textbf {\bibinfo
  {volume} {49}},\ \bibinfo {pages} {16349} (\bibinfo {year}
  {1994})}\BibitemShut {NoStop}%
\bibitem [{\citenamefont {Jones}(1999)}]{Jones1999}%
  \BibitemOpen
  \bibfield  {author} {\bibinfo {author} {\bibfnamefont {R.~O.}\ \bibnamefont
  {Jones}},\ }\href {\doibase 10.1063/1.478414} {\bibfield  {journal} {\bibinfo
   {journal} {The Journal of Chemical Physics}\ }\textbf {\bibinfo {volume}
  {110}},\ \bibinfo {pages} {5189} (\bibinfo {year} {1999})}\BibitemShut
  {NoStop}%
\bibitem [{\citenamefont {Borovikov}\ and\ \citenamefont
  {Zangwill}(2009)}]{Borovikov2009}%
  \BibitemOpen
  \bibfield  {author} {\bibinfo {author} {\bibfnamefont {V.}~\bibnamefont
  {Borovikov}}\ and\ \bibinfo {author} {\bibfnamefont {A.}~\bibnamefont
  {Zangwill}},\ }\href {\doibase 10.1103/PhysRevB.80.121406} {\bibfield
  {journal} {\bibinfo  {journal} {Physical Review B}\ }\textbf {\bibinfo
  {volume} {80}},\ \bibinfo {pages} {121406} (\bibinfo {year} {2009})},\
  \Eprint {http://arxiv.org/abs/0907.1606} {arXiv:0907.1606} \BibitemShut
  {NoStop}%
\bibitem [{\citenamefont {Jacobson}\ \emph {et~al.}(2012)\citenamefont
  {Jacobson}, \citenamefont {St{\"{o}}ger}, \citenamefont {Garhofer},
  \citenamefont {Parkinson}, \citenamefont {Schmid}, \citenamefont {Caudillo},
  \citenamefont {Mittendorfer}, \citenamefont {Redinger},\ and\ \citenamefont
  {Diebold}}]{Jacobson2012}%
  \BibitemOpen
  \bibfield  {author} {\bibinfo {author} {\bibfnamefont {P.}~\bibnamefont
  {Jacobson}}, \bibinfo {author} {\bibfnamefont {B.}~\bibnamefont
  {St{\"{o}}ger}}, \bibinfo {author} {\bibfnamefont {A.}~\bibnamefont
  {Garhofer}}, \bibinfo {author} {\bibfnamefont {G.~S.}\ \bibnamefont
  {Parkinson}}, \bibinfo {author} {\bibfnamefont {M.}~\bibnamefont {Schmid}},
  \bibinfo {author} {\bibfnamefont {R.}~\bibnamefont {Caudillo}}, \bibinfo
  {author} {\bibfnamefont {F.}~\bibnamefont {Mittendorfer}}, \bibinfo {author}
  {\bibfnamefont {J.}~\bibnamefont {Redinger}}, \ and\ \bibinfo {author}
  {\bibfnamefont {U.}~\bibnamefont {Diebold}},\ }\href {\doibase
  10.1021/jz2015007} {\bibfield  {journal} {\bibinfo  {journal} {The Journal of
  Physical Chemistry Letters}\ }\textbf {\bibinfo {volume} {3}},\ \bibinfo
  {pages} {136} (\bibinfo {year} {2012})}\BibitemShut {NoStop}%
\bibitem [{\citenamefont {Dojima}\ \emph {et~al.}(2016)\citenamefont {Dojima},
  \citenamefont {Kutsuma}, \citenamefont {Ashida},\ and\ \citenamefont
  {Kaneko}}]{dojima2016kinetically}%
  \BibitemOpen
  \bibfield  {author} {\bibinfo {author} {\bibfnamefont {D.}~\bibnamefont
  {Dojima}}, \bibinfo {author} {\bibfnamefont {Y.}~\bibnamefont {Kutsuma}},
  \bibinfo {author} {\bibfnamefont {K.}~\bibnamefont {Ashida}}, \ and\ \bibinfo
  {author} {\bibfnamefont {T.}~\bibnamefont {Kaneko}},\ }in\ \href@noop {}
  {\emph {\bibinfo {booktitle} {The European Conference on Silicon Carbide and
  Related Materials}}}\ (\bibinfo {year} {2016})\ pp.\ \bibinfo {pages}
  {609--610}\BibitemShut {NoStop}%
\end{thebibliography}%

\end{document}